\documentclass[fleqn,3p,onecolumn,11pt,nofootinbib,showkeys,showpacs]{revtex4-1}

\newcommand{\versionnumber}{v1.00}

\renewcommand{\today}{13.10.2013 (\versionnumber)}
\usepackage[small,justification=centerlast]{caption}
\usepackage{graphicx}
\usepackage{bm}
\usepackage{mathrsfs}
\usepackage[latin1]{inputenc}
\usepackage{stmaryrd}
\usepackage{array}
\usepackage{psfrag}
\usepackage{dsfont}
\usepackage{epsfig}
\usepackage{titletoc}
\usepackage{float}   %  for figures
\usepackage{wrapfig} %  for figures
\usepackage{amsmath}
\usepackage[latin1]{inputenc}
\usepackage{amsmath}\usepackage{amssymb,stmaryrd}
\usepackage{mathrsfs}
\usepackage{array}
\usepackage{graphicx}
\usepackage{psfrag}
\usepackage{dsfont}
\usepackage{epsfig}
\usepackage{cancel}
\usepackage[dvips]{feynmp}
\usepackage{upgreek}
\usepackage{subfig}
\usepackage{tabularx}
\usepackage{color}
\usepackage{units}

\usepackage{textfit} %\scaletoheight{0.05cm}{very small text}
\usepackage{hyperref}

\renewcommand{\eqref}[1]{\mbox{Eq.~(\ref{#1})}}
\newcommand{\tabref}[1]{\mbox{Tab.~\ref{#1}}}
\newcommand{\figref}[1]{\mbox{Fig.~\ref{#1}}}
\newcommand{\secref}[1]{\mbox{Sec.~\ref{#1}}}

\begin{document}

\title{Quantum field theoretic properties of Lorentz-violating operators of nonrenormalizable dimension in the photon sector}

\author{M. Schreck} \email{mschreck@indiana.edu}
\affiliation{Indiana University Center for Spacetime Symmetries, Indiana University, Bloomington, Indiana 47405-7105}

\begin{abstract}

In the context of the nonminimal Standard-Model Extension a special subset of the \textit{CPT}-even higher-dimensional operators in
the photon sector is discussed from a quantum-field theoretical point of view. The modified dispersion laws, photon polarization
vectors plus the gauge field propagator are obtained and their properties are analyzed. It is demonstrated that for certain sectors
of the modified theory a puzzle arises for the optical theorem at tree-level. This is followed by a discussion of how it can be
interpreted and resolved at first order Lorentz violation. Furthermore the commutator of two gauge fields that are evaluated at different
spacetime points is obtained and discussed. The structure of the theory is shown to resemble the structure of the modification based on
the corresponding dimension-4 operator. However some properties are altered due to the nonrenormalizable nature of the theory considered.
The results provide more insight into the characteristics of Lorentz-violating quantum field theories that rest upon contributions of
nonrenormalizable dimension.

\end{abstract}
%11.15.-q gauge field theory

\maketitle

\newpage%%tmp
\setcounter{equation}{0}
\setcounter{section}{0}
\renewcommand{\theequation}{\arabic{section}.\arabic{equation}}

\section{Introduction}

Over the past 15 years the study of Lorentz invariance violation both in theory and experiment has become an important field. The foundations
were laid by the seminal papers \cite{Kostelecky:1988zi,Kostelecky:1991ak,Kostelecky:1994rn} in which it was shown that a violation of Lorentz
symmetry can emerge in certain scenarios of string theory. In addition, Lorentz violation may arise in many other interesting contexts such as
a nontrivial structure of spacetime (spacetime foam) \cite{Wheeler:1957mu,Hawking:1979zw,Bernadotte:2006ya,Klinkhamer:2003ec}, noncommutative
field theories \cite{Carroll:2001ws}, loop quantum gravity \cite{Gambini:1998it,Bojowald:2004bb}, and quantum field theories on spacetimes with
a nontrivial topological structure \cite{Klinkhamer:1999zh,Klinkhamer:1998fa}.

In principle it is assumed that a low-energy effective description of quantum gravity phenomena can be considered as an expansion in energy over
a mass scale, which is probably related to the Planck scale. The leading-order term in such an expansion comprises the ordinary Standard Model of
elementary particle physics plus General Relativity. The next-to-leading order term is the minimal Standard-Model Extension (SME) \cite{Colladay:1998fq},
which is a framework to study and test Lorentz violation at energies much smaller than the Planck scale. The minimal SME includes all Lorentz-violating
operators that are invariant under the gauge group of the Standard Model plus power-counting renormalizable. Since gravity itself is nonrenormalizable
one may expect higher-order terms in the expansion to be made up of operators of nonrenormalizable dimension. These are included in the nonminimal SME.

A special sector of the nonrenormalizable SME forms the basis of the current paper.
A necessary (but not sufficient) criterion for a renormalizable, interacting quantum field theory is that it only contains products of field operators
that have a mass dimension of four or less. Operators with mass dimension of at least five are called higher-dimensional. In the
early days of the development and understanding of renormalization many theorists considered such quantum field theories with antipathy.
However, this point of view has changed. Nowadays nonrenormalizable quantum field theories are powerful tools in both high-energy and condensed
matter physics. Such theories have significance as long as they are considered as an effective theory only valid within a certain energy
range \cite{Georgi:1994qn,Pich:1998xt}.

There are many examples for effective theories: Fermi's theory of the weak interaction \cite{Fermi:1934sk}, Euler-Heisenberg theory \cite{Heisenberg:1935qt},
chiral perturbation theory, heavy quark effective theory, etc. (for the latter two cf. the review \cite{Pich:1998xt}). Even the Standard Model
of elementary particle physics can be considered as an effective (though
renormalizable) theory and practically all condensed matter theories are effective ones. To illustrate why these are successful we
consider Euler-Heisenberg theory as an example for an effective theory in the photon sector.

Let us assume that we only knew about classical physics, e.g., the Lagrange density of classical electrodynamics, which is proportional to
a bilinear combination of electromagnetic field strength tensors, $F^2$, and leads to Maxwell's equations. Based on Lorentz invariance, nothing
would forbid us to add higher-dimensional terms to this Lagrange density. Two of the possible terms are $F^4$ and $(F\widetilde{F})^2$ with the dual
field strength tensor $\widetilde{F}$, where each of them is multiplied with an unknown coefficient of mass dimension $-4$. These would then lead
to nonlinear versions of Maxwell's equations describing light-by-light scattering. The latter is certainly not a classical but a quantum
theoretical phenomenon. Hence even without any knowledge of quantum theory we could gain an understanding of it if we were able to determine
the unknown coefficients by experiment.
However, since we know about quantum theory, these coefficients can be calculated in perturbative quantum electrodynamics (QED) and they are
inversely proportional to the  fourth power of the electron mass. Although Euler-Heisenberg theory is nonrenormalizable, it gives a good
description for quantum effects in electrodynamics as long as the the photon energy is much smaller than the electron mass. The
nonrenormalizable nature of the theory is revealed
since the cut-off dependence of scattering quantities cannot be removed by a redefinition of the theory parameters. When the cut-off reaches
the electron mass the higher-dimensional terms can be as large as the renormalizable ones and the validity of the theory breaks down.

The example given shall demonstrate how useful nonrenormalizable field theories can still be. This was certainly one motivation for
Kosteleck\'{y} and Mewes to include higher-dimensional operators in the photon \cite{Kostelecky:2009zp}, neutrino \cite{Kostelecky:2011gq},
and the fermion sector \cite{Kostelecky:2013rta} of the minimal SME, which leads us to the nonminimal SME. Since Lorentz violation is
supposed to originate from physics at the Planck scale, these terms of nonrenormalizable dimension are probably suppressed by the Planck mass. 
The theory is applicable as long as particle momenta do not lie in the order of magnitude of this scale.

The current paper shall provide a better insight on the quantum field theoretical properties of such Lorentz-violating
nonrenormalizable theories. Here the focus is on the nonminimal, \textit{CPT}-even photon sector whose terms are classified in
\cite{Kostelecky:2009zp}. In \secref{sec:nonminimal-cpt-even-sme} the action of the theory considered will be introduced and its
general properties will be discussed. The framework is then restricted to a particular subset of Lorentz-violating parameters.
Based upon this modification, the modified dispersion relations of electromagnetic waves are obtained and investigated in
\secref{sec:dispersion-relations-classical-theory}, which is followed by the calculation of the polarization vectors and the photon
propagator in \secref{sec:polarization-vector-propagator}. Using these results the optical theorem at tree-level will be checked
in \secref{sec:discussion-of-optical-theorem}. It is demonstrated that a puzzle arises for certain sectors of the theory and how to
resolve it at leading order Lorentz violation. Section \ref{sec:gauge-field-commutator} is dedicated to studying the properties
of the gauge potential commutator of the theory with the goal to get some understanding of its causal structure. The results
are concluded in \secref{sec:conclusions}. Calculational details are relegated to Apps. \ref{sec:normalization-polarization-vectors}
and \ref{sec:gauge-field-commutator-calculation}. Natural units with $\hbar=c=1$ will be used throughout the paper unless stated
otherwise.

\section{\textit{CPT}-even photon sector of the nonminimal Standard-Model Extension}
\label{sec:nonminimal-cpt-even-sme}

Within this paper modified Maxwell theory shall be considered, which is the \textit{CPT}-even modification of the SME photon sector.
Including the dimension-4 and all higher-dimensional contributions this theory is defined by the following action:
\begin{subequations}
\label{eq:action-modified-maxwell-theory}
\begin{eqnarray}
S_{\mathrm{modMax}}&=&\int_{\mathbb{R}^4}\mathrm{d}^4x\,
\mathcal{L}_\text{modMax}(x)\,, \\[2mm]
\label{eq:lagrange-density-modmax}
\mathcal{L}_\text{modMax}(x)&=& -\frac{1}{4}
\eta^{\mu\rho}\eta^{\nu\sigma}F_{\mu\nu}(x)F_{\rho\sigma}(x)
-\frac{1}{4}(\widehat{k}_F)^{\mu\nu\varrho\sigma}F_{\mu\nu}(x)F_{\varrho\sigma}(x)\,, \\[2mm]
\label{eq:L-modified-maxwell-theory}
(\widehat{k}_F)^{\mu\nu\varrho\sigma}&=&\sum_{\substack{d=\text{even} \\ d\geq 4}} (k_F^{(d)})^{\mu\nu\varrho\sigma\alpha_1\dots\alpha_{(d-4)}}\partial_{\alpha_1}\dots \partial_{\alpha_{(d-4)}}\,.
\end{eqnarray}
\end{subequations}
The action is written in terms of the electromagnetic field strength tensor $F_{\mu\nu}(x)\equiv\partial_{\mu}A_{\nu}(x)-\partial_{\nu}A_{\mu}(x)$
of the \textit{U}(1) gauge field $A_{\mu}(x)$. The fields are defined on Minkowski spacetime with coordinates
$(x^\mu)=(x^0,\boldsymbol{x})=(c\,t,x^1,x^2,x^3)$ and metric $(g_{\mu\nu}(x))=(\eta_{\mu\nu}) \equiv \text{diag}\,(1,-1,-1,-1)$.
The Lagrange density of \eqref{eq:lagrange-density-modmax} is decomposed into the standard Maxwell term and the Lorentz-violating modification.
The latter involves the background coefficients $(\widehat{k}_F)^{\mu\nu\varrho\sigma}$, which transform covariantly with respect to observer
Lorentz transformations and are fixed with respect to particle Lorentz transformations.

Besides the dimension-4 modified Maxwell term (cf. Refs.~\cite{ChadhaNielsen1983,Colladay:1998fq,KosteleckyMewes2002}) the background
coefficients involve all higher-dimensional operators with even operator dimension $d>4$. The terms of nonrenormalizable dimension are characterized
by additional derivatives. Each increase of the operator dimension by two involves two additional derivatives that are contracted
with an appropriate Lorentz-violating coefficient with two additional indices. The mass dimension of this coefficient is decreased
by two to compensate the mass dimensions of the derivatives. Transforming these coefficients to momentum space with the four-momentum
$k_{\alpha}$ yields:
\begin{equation}
\label{eq:dimension-d-operator}
(k_F^{(d)})^{\mu\nu\varrho\sigma\alpha_1\alpha_2}\partial_{\alpha_1}\dots \partial_{\alpha_{(d-4)}} \xrightarrow[\text{momentum space}]{\partial_{\alpha} \mapsto \mathrm{i}k_{\alpha}} (-1)^{d/2-2}(k_F^{(d)})^{\mu\nu\varrho\sigma\alpha_1\alpha_2}k_{\alpha_1}\dots k_{\alpha_{(d-4)}}\,.
\end{equation}
Hence the scales of the individual contributions in the expansion of $(\widehat{k}_F)^{\mu\nu\varrho\sigma}$ can be made more transparent with
the following symbolic notation:
\begin{equation}
\label{eq:dimensional-expansion}
|(\widehat{k}_F)^{\mu\nu\varrho\sigma}|=|\kappa^{\mu\nu\varrho\sigma}|\left(1+\zeta^{(2)}k^2+\zeta^{(4)}k^4+\dots\right)\,,
\end{equation}
where $|\bullet|$ denotes the order of magnitude of the Lorentz-violating coefficients and $\kappa^{\mu\nu\varrho\sigma}$ are the coefficients 
associated with the dimension-4 operator, $(k_F^{(4)})^{\mu\nu\varrho\sigma}\equiv \kappa^{\mu\nu\varrho\sigma}$. The quantities $\zeta^{(d-4)}$
(for even $d>4$) have mass dimension $4-d$ and the variable $k$ denotes a momentum scale. A necessary condition for this expansion to be 
well-defined is that $\zeta^{(d-4)}k^{d-4}\ll 1$. The leading dimension-4 operator is often called ``marginal'' and the subleading terms are
named ``irrelevant.''

Equation (\ref{eq:dimensional-expansion}) means that the marginal operator is dominant as long as $\zeta^{(d-4)}k^{d-4}\ll 1$. However, the
larger the momentum of a photon the more important are the higher-dimensional operators, which is why the expression ``irrelevant'' can be
misleading in this case. Keep in mind that the validity of the effective theory breaks down when $k$ approaches an order of magnitude such
that $\zeta^{(d-4)}k^{d-4}=\mathcal{O}(1)$. Then all terms in the expansion above become equally important and it is not supposed to converge
any more.

The properties of the quantum field theory based on the marginal operator and all the irrelevant ones set to zero have been investigated
in the series of papers \cite{Casana-etal2009,Casana-etal2010,Klinkhamer:2010zs,Schreck:2011ai,Schreck:2013gma}. The current goal is to
understand the implications of including some of the higher-dimensional ones. To keep the calculations feasible we restrict the analysis
to a particular subset of operators. It is natural to consider the first of the higher-dimensional operators that has mass dimension six
where all remaining ones are set to zero:
\begin{equation}
\label{eq:dimension-six-operator}
(\widehat{k}_F)^{\mu\nu\varrho\sigma}=(k_F^{(6)})^{\mu\nu\varrho\sigma\alpha_1\alpha_2}\partial_{\alpha_1}\partial_{\alpha_2} \xrightarrow[\text{momentum space}]{\partial_{\alpha} \mapsto -\mathrm{i}k_{\alpha}} -(k_F^{(6)})^{\mu\nu\varrho\sigma\alpha_1\alpha_2}k_{\alpha_1}k_{\alpha_2}\,.
\end{equation}
There is a generalization of the nonbirefringent \textit{ansatz} of the dimension-4 operator \cite{BaileyKostelecky2004,Altschul:2006zz}.
In particular for the dimension-6 operator it is given by \cite{Kostelecky:2009zp}
\begin{align}
\label{eq:nonbirefringent-ansatz}
(k_F^{(6)})^{\mu\nu\varrho\sigma\alpha_1\alpha_2}k_{\alpha_1}k_{\alpha_2}&=\frac{1}{2}\left[\eta^{\mu\varrho}(c_{\scriptscriptstyle{F}})^{\nu\sigma\alpha_1\alpha_2}-\eta^{\nu\varrho}(c_{\scriptscriptstyle{F}})^{\mu\sigma\alpha_1\alpha_2}\right. \notag \\
&\phantom{{}={}\frac{1}{2}\Big[}\left.+\,\eta^{\nu\sigma}(c_{\scriptscriptstyle{F}})^{\mu\varrho\alpha_1\alpha_2}-\eta^{\mu\sigma}(c_{\scriptscriptstyle{F}})^{\nu\varrho\alpha_1\alpha_2}\right]k_{\alpha_1}k_{\alpha_2}\,,
\end{align}
with the Minkowski metric $\eta^{\mu\nu}$ and a four-tensor $(c_{\scriptscriptstyle{F}})^{\mu\nu\varrho\sigma}$.
Now we want to restrict ourselves to the case that is equivalent to the isotropic sector of the dimension-4 operator. The corresponding
$(c_{\scriptscriptstyle{F}})^{\mu\nu\alpha_1\alpha_2}k_{\alpha_1}k_{\alpha_2}$ is a $(4\times 4)$-matrix that has the same form as the
respective matrix \cite{Kaufhold:2007qd,Casana-etal2010} of the isotropic sector mentioned. The isotropic dimensionless coefficient
$\widetilde{\kappa}_{\mathrm{tr}}$ of the dimension-4 operator is replaced by the combination
$\kappa_{\mathrm{tr}-}^{\alpha_1\alpha_2}k_{\alpha_1}k_{\alpha_2}$ that now appears in each entry of the matrix:
\begin{equation}
\label{eq:dimension-6-isotropic-sector-minus}
(c_{\scriptscriptstyle{F}})^{\mu\nu\alpha_1\alpha_2}k_{\alpha_1}k_{\alpha_2}=\left(2\xi^{\mu}\xi^{\nu}-\frac{\xi^2}{2}\eta^{\mu\nu}\right)\kappa_{\mathrm{tr}-}^{\alpha_1\alpha_2}k_{\alpha_1}k_{\alpha_2}=\frac{1}{2}\mathrm{diag}(3,1,1,1)^{\mu\nu}\kappa_{\mathrm{tr}-}^{\alpha_1\alpha_2}k_{\alpha_1}k_{\alpha_2}\,,
\end{equation}
with the timelike four-vector $(\xi^{\mu})=(1,0,0,0)^T$.
All remaining Lorentz-violating coefficients are assumed to vanish. The coefficients $\kappa_{\mathrm{tr}-}^{\mu\nu}$ have mass dimension $-2$.
The minus sign in their index was added to distinguish them from the other set of isotropic parameters, $\kappa_{\mathrm{tr}+}$. In
the context of the coefficients $\kappa^{\mu\nu\varrho\sigma}$, which are contained in the dimension-4 operator, $\kappa_{\mathrm{tr}+}$
corresponds to the double trace $\kappa^{\mu\nu}_{\phantom{\mu\nu}\mu\nu}$. Since the latter can be removed by a field
redefinition \cite{Colladay:1998fq}, $\kappa_{\mathrm{tr}+}$ does not describe any physics when the theory is restricted to the marginal
operator. On the contrary, it can lead to physical effects for the higher-dimensional operators~\cite{Kostelecky:2009zp}, but it is
discarded here for simplicity.

In what follows, the Lorentz-violating nonminimal SME sector characterized by Eqs. (\ref{eq:action-modified-maxwell-theory}),
(\ref{eq:dimension-six-operator}), (\ref{eq:nonbirefringent-ansatz}), and (\ref{eq:dimension-6-isotropic-sector-minus}) shall be
studied. Instead of only the coefficient $\widetilde{\kappa}_{\mathrm{tr}}$ for the marginal operator there now exists a $(4\times 4)$-matrix
$(\kappa_{\mathrm{tr}-}^{\alpha_1\alpha_2})$ that makes up this sector. Since this matrix is combined with the symmetric two-tensor
$k_{\alpha_1}k_{\alpha_2}$, its antisymmetric part can be discarded. Therefore, $(\kappa_{\mathrm{tr}-}^{\alpha_1\alpha_2})$ is assumed to be
completely symmetric. Furthermore, antisymmetrization on any triple of indices of $(k_F^{(6)})^{\mu\nu\varrho\sigma\alpha_1\alpha_2}$
produces zero \cite{Kostelecky:2009zp}, which reduces the number of independent coefficients further. However these additional restrictions are
not important for the current paper, and they are not considered.

Certain properties of the modified electrodynamics cannot be investigated without a coupling to matter. Therefore, the modified free theory
is minimally coupled to a standard Dirac theory of spin-1/2 fermions with charge $e$ and mass $m$. This results in a Lorentz-violating
extended QED that is defined by the following action:
\begin{equation}\label{eq:action-birefringent-qed} \hspace*{0mm}
S_\text{modQED}^{\text{isotropic } d=6}\big[\kappa_{\mathrm{tr}-}^{\mu\nu},e,m\big] =
S_\text{modMax}^{\text{isotropic } d=6}\big[\kappa_{\mathrm{tr}-}^{\mu\nu}\big] +
S^\text{}_\text{Dirac}\big[e,m\big]\,.
\end{equation}
The Lorentz-violating \textit{CPT}-even modification of the gauge field $A_\mu(x)$ is given by Eqs. (\ref{eq:action-modified-maxwell-theory}),
(\ref{eq:dimension-six-operator}), (\ref{eq:nonbirefringent-ansatz}), and (\ref{eq:dimension-6-isotropic-sector-minus}). The standard Dirac
term for the spinor field $\psi(x)$ reads
\begin{subequations}
\label{eq:standDirac-action}
\begin{align}
S^\text{ }_\text{Dirac}\big[e,M\big] &=
\int_{\mathbb{R}^4} \mathrm{d}^4 x \; \overline\psi(x) \left[
\gamma^\mu \left(\frac{\mathrm{i}}{2}\,\overleftrightarrow{\partial_\mu} -e A_\mu(x) \right) -m\right] \psi(x)\,, \\[2ex]
A\overleftrightarrow{\partial_{\mu}}B&\equiv A\partial_{\mu}B-(\partial_{\mu}A)B\,.
\end{align}
\end{subequations}
The latter action contains the standard Dirac matrices $\gamma^\mu$ with the Clifford algebra $\{\gamma^{\mu},\gamma^{\nu}\}=2\eta^{\mu\nu}$
and it is written such that the respective Lagrange density is Hermitian.

\section{Dispersion relations}
\label{sec:dispersion-relations-classical-theory}
\setcounter{equation}{0}

The first step is to obtain the modified dispersion relations of electromagnetic waves. The field equations for the theory based on the higher-dimensional
operators have the same form as those of the dimension-4 \textit{CPT}-even extension \cite{Colladay:1998fq,KosteleckyMewes2002,BaileyKostelecky2004}.
They are given by:
\begin{equation}
\label{eq:field-equations-modified-maxwell-theory}
M^{\mu\nu}A_{\nu}=0 \,,\quad
M^{\mu\nu}\equiv
k^{\rho}k_{\rho}\,\eta^{\mu\nu}-k^{\mu}k^{\nu}
-2\,(\widehat{k}_F)^{\mu\rho\sigma\nu}\,k_{\rho}k_{\sigma}\,,
\end{equation}
with the four-momentum $k^{\mu}$. The modified dispersion relations are the conditions that have to be fulfilled by the
four-momentum such that \eqref{eq:field-equations-modified-maxwell-theory} has nontrivial solutions for $A_{\nu}$. They
follow from $\det(M)=0$ with the matrix $M$ given in \eqref{eq:field-equations-modified-maxwell-theory}.
To obtain the dispersion laws it is convenient to divide the matrix $(\kappa_{\mathrm{tr}-}^{\mu\nu})$ into three parts:
\begin{equation}
(\kappa_{\mathrm{tr}-}^{\mu\nu})=\left(\begin{array}{c|ccc}
\kappa_{\mathrm{tr}-}^{00} & \kappa_{\mathrm{tr}-}^{01} & \kappa_{\mathrm{tr}-}^{02} & \kappa_{\mathrm{tr}-}^{03} \\
\hline
\kappa_{\mathrm{tr}-}^{01} & \kappa_{\mathrm{tr}-}^{11} & \kappa_{\mathrm{tr}-}^{12} & \kappa_{\mathrm{tr}-}^{13} \\
\kappa_{\mathrm{tr}-}^{02} & \kappa_{\mathrm{tr}-}^{12} & \kappa_{\mathrm{tr}-}^{22} & \kappa_{\mathrm{tr}-}^{23} \\
\kappa_{\mathrm{tr}-}^{03} & \kappa_{\mathrm{tr}-}^{13} & \kappa_{\mathrm{tr}-}^{23} & \kappa_{\mathrm{tr}-}^{33} \\
\end{array}
\right)\,.
\end{equation}
The first contains the single coefficient $\kappa_{\mathrm{tr}-}^{00}$ that appears together with two time derivatives. For
this reason it will be denoted as the ``temporal part.'' The second sector is characterized by the three mixed coefficients
$\kappa_{\mathrm{tr}-}^{0i}$ for $i=1$, 2, 3 that are combined with one time and one spatial derivative. Hence we call it the
``mixed part.'' Finally, the third sector is made up of the six spatial coefficients $\kappa_{\mathrm{tr}-}^{ij}$
for $i\leq j$, $j=1$, 2, 3, whereby it is named the ``spatial part.'' The following investigations will be performed for
these three sectors separately.

\subsection{Temporal part}
\label{sec:dispersion-relations-purely-temporal}

In this case all coefficients are set to zero expect of $\kappa_{\mathrm{tr}-}^{00}$. The determinant of $M$ involves a
biquadratic polynomial whose solutions with respect to $k^0$ correspond to the two physical dispersion relations that are
given as follows:
\begin{equation}
\label{eq:dispersion-law-purely-spatial}
\omega_{1,2}(\mathbf{k})=\frac{1}{\sqrt{2\kappa_{\mathrm{tr}-}^{00}}}\sqrt{1-\kappa_{\mathrm{tr}-}^{00}\,\mathbf{k}^2\mp \sqrt{1-\kappa_{\mathrm{tr}-}^{00}\,\mathbf{k}^2(6-\kappa_{\mathrm{tr}-}^{00}\,\mathbf{k}^2)}}\,.
\end{equation}
Note that $\kappa_{\mathrm{tr}-}^{00}$ has mass dimension $-2$ whereby it always occurs in combination with $\mathbf{k}^2$ to produce a dimensionless
quantity. Since the second square root appears with two different signs there are two distinct dispersion relations with different phase velocities.
To get a better insight in this issue the following expansions for both dispersion
laws are given for $\kappa_{\mathrm{tr}-}^{00}\mathbf{k}^2\ll 1$:
\begin{subequations}
\begin{align}
\omega_1(\mathbf{k})&=|\mathbf{k}|+(\kappa^{00}_{\mathrm{tr}-})|\mathbf{k}|^3+\frac{5}{2}(\kappa^{00}_{\mathrm{tr}-})^2|\mathbf{k}|^5+\dots\,, \\[2ex]
\label{eq:dispersion-isotropic-spurious}
\omega_2(\mathbf{k})&=\frac{1}{\sqrt{\kappa^{00}_{\mathrm{tr}-}}}-\sqrt{\kappa^{00}_{\mathrm{tr}-}}\mathbf{k}^2-\frac{3}{2}(\kappa^{00}_{\mathrm{tr}-})^{3/2}\mathbf{k}^4+\dots\,.
\end{align}
\end{subequations}
The first dispersion law is a perturbation of the standard dispersion relation $\omega(\mathbf{k})=|\mathbf{k}|$, whereas the second does not have
an existing limit for $\kappa^{00}_{\mathrm{tr}-}\mapsto 0$. Such a behavior does not occur for the sectors of the dimension-4 operators that were
considered in \cite{Casana-etal2009,Casana-etal2010,Klinkhamer:2010zs,Schreck:2011ai,Schreck:2013gma}. The existence of $\omega_2$ traces back
to treatment of the Lorentz-violating extension as an effective field theory. According to~\cite{Kostelecky:2009zp} such dispersion laws are neglected
since they do not arise as a small perturbation from the standard relations. They must be considered as Planck scale effects. If $\kappa^{00}_{\mathrm{tr}-}$
is indeed nonzero in nature, modes that are associated with $\omega_2$ may become especially important if the momentum $|\mathbf{k}|$ approaches the
Planck scale. However keep in mind that we are dealing with an effective field theory whose applicability is expected to break down for momenta
in the vicinity of the Planck scale (see also the discussion at the end of Sec.~(IIc) in \cite{Kostelecky:2009zp}).

For the reasons mentioned, modified dispersion relations that are a perturbation of the standard one will be called ``perturbed'' and the others,
which are not a perturbation, will be referred to as ``spurious.'' Note that the spurious dispersion relation of \eqref{eq:dispersion-isotropic-spurious} 
is, in fact, associated
with one of the transverse, propagating modes. This can be shown with the modified Coulomb and Amp\`{e}re law according to \cite{Colladay:1998fq}.
Using the latter procedure unphysical dispersion laws being associated to the scalar and longitudinal mode can be identified and discarded. This
procedure cannot be applied to remove the spurious dispersion law, though.

\subsection{Mixed part}
\label{subsec:mixed-part}

For the mixed part the determinant of $M$ is more complicated and contains a third order polynomial in $k_0$. Note that this is the first of all
sectors of modified Maxwell theory studied so far where the physical dispersion laws result from a polynomial of this degree. This renders the
calculation of the dispersion relations more complicated in comparison to the aforementioned sectors. With the transformation
\begin{equation}
\label{eq:transformation-cardano}
k_0=k_0'-\frac{a}{3}\,,\quad a=a(\mathbf{k})=\frac{1}{2\kappa_{\mathrm{tr}-}^{0i}k^i}\,,
\end{equation}
and an additional multiplication with $-a$ the third order polynomial can be recast in the form $k_0'^3+pk_0'+2q$ with
\begin{subequations}
\begin{align}
\label{eq:definition-cardano-p-q}
p=p(\mathbf{k})&=\mathbf{k}^2-\frac{1}{12(\kappa_{\mathrm{tr}-}^{0i}k^i)^2}\,,\quad
q=q(\mathbf{k})=\frac{1-72\mathbf{k}^2(\kappa_{\mathrm{tr}-}^{0i}k^i)^2}{216(\kappa_{\mathrm{tr}-}^{0i}k^i)^3}\,, \\[2ex]
d=d(\mathbf{k})&=q^2+\left(\frac{p}{3}\right)^3\,,
\end{align}
\end{subequations}
where a summation over $i=1$, 2, 3 is understood and $d$ will be needed below.
Two of the three zeros of this polynomial with respect to $k_0'$ that are transformed back to $k_0$ via \eqref{eq:transformation-cardano}
correspond to the dispersion laws of the transverse degrees of freedom. They are given by:
\begin{subequations}
\label{eq:dispersion-laws-transverse}
\begin{align}
\label{eq:dispersion-laws-transverse-part1}
\omega_1(\mathbf{k})&=\sqrt{-\frac{4}{3}p}\cos\left[\frac{1}{3}\arccos\left(-q\sqrt{-\frac{27}{p^3}}\right)\right]-\frac{a}{3}\,, \\[2ex]
\label{eq:dispersion-laws-transverse-part2}
\omega_2(\mathbf{k})&=-\sqrt{-\frac{4}{3}p}\cos\left[\frac{1}{3}\arccos\left(-q\sqrt{-\frac{27}{p^3}}\right)+\frac{\pi}{3}\right]-\frac{a}{3}\,,
\end{align}
\end{subequations}
with $p$ and $q$ defined by \eqref{eq:definition-cardano-p-q}. Dependent on the sign of the functions $a$, $q$, $p$, and $d$ one of
these solutions is a perturbation of the standard dispersion law and the other one is a spurious dispersion relation similar to
\eqref{eq:dispersion-isotropic-spurious}. For $a>0$ and $q>0$ (where $p$ is negative for this choice) or $a>0$, $q<0$, and $p<0$
the only dispersion relation being a perturbation of the standard dispersion law is $\omega_1(\mathbf{k})$, which can then be 
rewritten as follows:
\begin{subequations}
\label{eq:dispersion-laws-transverse-1-total}
\begin{align}
\label{eq:dispersion-laws-transverse-1}
\omega_1(\mathbf{k})&=\sqrt[3]{u}+\sqrt[3]{v}-\frac{a}{3}\,, \\[2ex]
u&=u(\mathbf{k})=-q+\sqrt{d}\,,\quad v=v(\mathbf{k})=-q-\sqrt{d}\,.
\end{align}
\end{subequations}
For $a>0$, $q<0$, and $p>0$ or $a<0$, $q<0$ (where $p$ is negative in this case) or $a<0$, $q>0$, and $d<0$ only $\omega_2(\mathbf{k})$
is such a perturbation, which is rearranged to give:
\begin{subequations}
\label{eq:dispersion-laws-transverse-2-total}
\begin{align}
\label{eq:dispersion-laws-transverse-2}
\omega_2(\mathbf{k})&=-\frac{\mathrm{sign}(q)}{2}\sqrt{\frac{4}{3}p}\left[f(x)^{1/6}-\frac{1}{f(x)^{1/6}}\right]-\frac{a}{3}\,, \\[2ex]
f(\psi)&=\psi+\sqrt{\psi+1}\sqrt{\psi-1}\,,\quad x=1+\frac{54q^2}{p^3}\,.
\end{align}
\end{subequations}
Both $\omega_1$ and $\omega_2$ have been recast, from which it can be shown that they both are real quantities. The cubic
roots in \eqref{eq:dispersion-laws-transverse-1} give opposite imaginary parts that cancel in the sum. The second dispersion
law of \eqref{eq:dispersion-laws-transverse-2} is manifestly real for $p>0$. For $p<0$ the real parts arising in each of the
two terms in the square brackets cancel, which gives a purely imaginary result. Combining it with the imaginary result from the
square root outside of the brackets leads to a real quantity. These properties are not directly evident from
Eqs.~(\ref{eq:dispersion-laws-transverse-part1}), (\ref{eq:dispersion-laws-transverse-part2}).

For $a<0$, $q>0$, and $d>0$ the polynomial does not have a real and positive zero. Hence this choice does not lead to a
modified dispersion law. Finally, \tabref{tab:choices-coefficients} shows the regions of $\kappa_{\mathrm{tr}-}^{0i}k^i$
that result in nonnegative $q$, $p$, and $d$.
\setlength{\extrarowheight}{3pt}
\begin{table}[t]
\centering
\begin{tabular}{m{3cm}m{7cm}m{7cm}m{7cm}}
\toprule
 & \multicolumn{1}{c}{$q\geq 0$} & \multicolumn{1}{c}{$p\geq 0$} & \multicolumn{1}{c}{$d\geq 0$} \\
\colrule
\multicolumn{1}{c}{$\kappa_{\mathrm{tr}-}^{0i}k^i\geq 0$} & \multicolumn{1}{c}{$\kappa_{\mathrm{tr}-}^{0i}k^i \leq (6\sqrt{2}|\mathbf{k}|)^{-1}$} & \multicolumn{1}{c}{$\kappa_{\mathrm{tr}-}^{0i}k^i\geq (2\sqrt{3}|\mathbf{k}|)^{-1}$} & \multicolumn{1}{c}{$\kappa_{\mathrm{tr}-}^{0i}k^i\geq \sqrt{5\sqrt{5}-11}(2\sqrt{2}|\mathbf{k}|)^{-1}$} \\
\multicolumn{1}{c}{$\kappa_{\mathrm{tr}-}^{0i}k^i< 0$}    & \multicolumn{1}{c}{$\kappa_{\mathrm{tr}-}^{0i}k^i \leq -(6\sqrt{2}|\mathbf{k}|)^{-1}$}   & \multicolumn{1}{c}{$\kappa_{\mathrm{tr}-}^{0i}k^i\leq -(2\sqrt{3}|\mathbf{k}|)^{-1}$} & \multicolumn{1}{c}{$\kappa_{\mathrm{tr}-}^{0i}k^i< -\sqrt{5\sqrt{5}-11}(2\sqrt{2}|\mathbf{k}|)^{-1}$} \\
\botrule
\end{tabular}
\caption{Regions of Lorentz-violating coefficients and momentum components leading to nonnegative $q$, $p$, and $d$, respectively.}
\label{tab:choices-coefficients}
\end{table}%

\subsection{Spatial part}

Finally, the case is considered where all coefficients are assumed to vanish if they contain at least one Lorentz index that is equal
to zero. The physical dispersion law then results from a polynomial of second degree and can be cast in the following form:
\begin{equation}
\label{eq:dispersion-relation-purely-spatial-part}
\omega(\mathbf{k})=\sqrt{\frac{1+(\kappa_{\mathrm{tr}-}^{ij})k^ik^j}{1-(\kappa_{\mathrm{tr}-}^{ij})k^ik^j}}\,|\mathbf{k}|\,.
\end{equation}
Note the similarity to the isotropic dispersion relation when considering a nonvanishing dimension-4 operator with
the isotropic coefficient $\widetilde{\kappa}_{\mathrm{tr}}$:
\begin{equation}
\omega(\mathbf{k})|^{\text{isotropic } d=4}_{\mathrm{modMax}}=\sqrt{\frac{1-\widetilde{\kappa}_{\mathrm{tr}}}{1+\widetilde{\kappa}_{\mathrm{tr}}}}\,|\mathbf{k}|\,.
\end{equation}
Contrary to the previous cases, there only exists a single dispersion law, which is a Lorentz-violating perturbation of the standard
dispersion relation. Hence for the spatial part of the dimension-6 operator the modified dispersion law has the same form as for
the isotropic dimension-4 operator with $\widetilde{\kappa}_{\mathrm{tr}}$ replaced by $-(\kappa_{\mathrm{tr}-}^{ij})k^ik^j$. The
physical dispersion relations found for the spatial and the mixed sector are no longer isotropic. This shows that higher-dimensional
operators of the isotropic \textit{CPT}-even modification of the photon sector can deliver anisotropic contributions to the
dispersion relations.

A last comment concerns the degeneracy of the transverse dispersion relations for all three sectors previously considered. Both the
perturbed and the spurious dispersion laws have a twofold degeneracy, i.e., they appear as a double zero of the determinant of the matrix
$M$ in \eqref{eq:field-equations-modified-maxwell-theory}. The latter degeneracy reflects the degeneracy of the quantum-mechanical photon
state. It is important that the degeneracy of the perturbed dispersion law is still twofold despite the occurrence of the spurious dispersion
relations. The reason is that the photon state degeneracy goes in many physical quantities, e.g., Planck's radiation law. Hence if it
was modified, Planck's law would change as well and the limit of vanishing Lorentz-violating coefficients would not describe the 
experimental measurements correctly.

\section{Polarization vectors and the propagator}
\label{sec:polarization-vector-propagator}

The \textit{CPT}-even Lorentz-violating modification considered is based on a higher-dimensional operator. Due to the additional derivatives that
are combined with this operator it is interesting to examine the quantum-field theoretic properties of the modification. To do so, the
modified polarization vectors and the propagator are needed and they are obtained as follows.

The propagator of a quantized field is an important object for studying the properties of the underlying quantum field theory. It is
the Green's function of the free-field equations of motion, i.e., the inverse of the differential operator that appears in these
equations. However, due to the infinite number of gauge degrees of freedom of the photon field an inverse does not exist as long as
no gauge fixing condition is imposed. For all cases of the \textit{CPT}-even dimension-4 operator considered so far
\cite{Casana-etal2009,Casana-etal2010,Klinkhamer:2010zs,Schreck:2011ai,Schreck:2013gma}, Feynman gauge
\cite{Veltman1994,ItzyksonZuber1980,PeskinSchroeder1995} has proven to be a convenient gauge choice. Hence, this gauge choice will be
implemented here as well. In practice this is done by adding the following gauge-fixing term to the Lagrange density:
\begin{equation}
\label{eq:gauge-fixing-feynman}
\mathcal{L}_{\mathrm{gf}}(x)=
-\frac{1}{2}\big(\partial_{\mu}\,A^{\mu}(x)\big)^2\,.
\end{equation}
By partial integration the action of the modified photon sector can be written as follows:
\begin{subequations}
\begin{equation}
S_{\mathrm{modMax}}^{\mathrm{Feynman}}=\frac{1}{2}\int_{\mathbb{R}^4} \mathrm{d}^4x\,A_{\mu}(G^{-1})^{\mu\nu}A_{\nu}\,,
\end{equation}
with the differential operator
\begin{equation}
\label{eq:differential-operator}
(G^{-1})^{\mu\nu}=\eta^{\mu\nu}\partial^2
-2\,(k_F)^{\mu\varrho\sigma\nu}\partial_{\varrho}\partial_{\sigma}\,.
\end{equation}
\end{subequations}
Transforming $(G^{-1})^{\mu\nu}$ to momentum space leads to
\begin{equation}
(\widehat{G}^{-1})^{\mu\nu}=-k^2\eta^{\mu\nu}+2(k_F)^{\mu\varrho\sigma\nu}k_{\varrho}k_{\sigma}\,.
\end{equation}
Now the system of equations $(\widehat{G}^{-1})^{\mu\nu}\widehat{G}_{\nu\lambda}=\mathrm{i}\,\delta^{\mu}_{\phantom{\mu}\lambda}$
must be solved where $\widehat{G}_{\mu\nu}$ is the propagator in momentum space. To understand the structure of the propagator it must
be expressed in a covariant form using the four-vectors and two-tensors that are available in this context. This is the metric tensor
$\eta^{\mu\nu}$, the four-vector $k^{\mu}$, and the preferred spacetime direction $\xi^{\mu}$, which appears in
\eqref{eq:dimension-6-isotropic-sector-minus}. The isotropic case based on the dimension-4 operator is characterized by $\xi^{\mu}$.
It is assumed that $\xi^{\mu}$ is the only preferred direction that plays a role for the dimension-6 operator as well. For this reason
the following \textit{ansatz} is made for the propagator:
\begin{equation}
\label{eq:propagator-ansatz}
\widehat{G}^{\mu\nu}\,
\big|^{\mathrm{Feynman}}
=-\mathrm{i}\Big\{\widehat{a}\,\eta^{\mu\nu}+\widehat{b}\,k^{\mu}k^{\nu}+\widehat{c}\,(k^{\mu}\xi^{\nu}+\xi^{\mu}k^{\nu})+\widehat{d}\,\xi^{\mu}\xi^{\nu}\Big\}\,\widehat{K}\,.
\end{equation}
The propagator coefficients $\widehat{a}$, $\dots$, $\widehat{d}$ plus the scalar part $\widehat{K}$ depend on the four-momentum
components: $\widehat{a}=\widehat{a}(k^0,\mathbf{k})$, etc. Since the propagator is a symmetric two-tensor, the \textit{ansatz} has to
respect this property. This can be checked to be the case in \eqref{eq:propagator-ansatz}. Due to this symmetry, from the
original 16 equations only ten have to be solved to obtain the propagator coefficients plus the scalar part.

The propagator of a quantum field describes its off-shell properties. To understand its on-shell characteristics the dispersion relations
are needed plus --- in case of the photon field --- the corresponding polarization vectors. The modified dispersion laws were already
obtained in \secref{sec:dispersion-relations-classical-theory}. The polarization vectors will be determined as follows. These form a
set of four four-vectors that is a basis of Minkowski spacetime. Only two of them describe physical, i.e., transverse photon
polarization states where the remaining two correspond to scalar and longitudinal degrees of freedom.  The transverse photon polarization
vectors are solutions of the field equations (\ref{eq:field-equations-modified-maxwell-theory}) with $k_0$ to be replaced by the
physical dispersion laws. For the temporal, the mixed, and the spatial sector they can be chosen as follows where $\mathbf{k}=(k_1,k_2,k_3)$
is a general three-momentum:
\begin{equation}
\label{eq:polarization-vectors}
\varepsilon^{(1)\,\mu}=\frac{1}{\sqrt{N^{(6)}}\,\sqrt{k_2^2+k_3^2}}\begin{pmatrix}
0 \\
0 \\
-k_3 \\
k_2 \\
\end{pmatrix}\,,\quad \varepsilon^{(2)\,\mu}=\frac{1}{\sqrt{N^{(6)}}\,|\mathbf{k}|}\begin{pmatrix}
0 \\
-\sqrt{k_2^2+k_3^2} \\
k_1k_2/\sqrt{k_2^2+k_3^2} \\
k_1k_3/\sqrt{k_2^2+k_3^2} \\
\end{pmatrix}\,,
\end{equation}
with a normalization $N^{(6)}$. The latter is an additional normalization, which is not related to the requirement that the scalar product
of a polarization vector with itself is equal to one. On the contrary, it has to be determined from the 00-component of the energy-momentum
tensor (given by Eq.~(36) in \cite{Colladay:1998fq}) whose expectation value must correspond to the modified physical photon dispersion law.
The procedure is described in App.~\ref{sec:normalization-polarization-vectors} in detail. Note that besides the appearance of $N^{(6)}$, the
polarization vectors are completely standard. It can be checked that they are orthogonal to each other and each is orthogonal to the momentum
three-vector. So they are interpreted as the physical transverse polarizations.

Equation (\ref{eq:polarization-vectors}) provides the polarization vectors of both the perturbed and the spurious dispersion 
relation. The reason is that the degeneracy of each dispersion law is still twofold as in the standard theory (cf. the discussion at the end of
\secref{sec:dispersion-relations-classical-theory}).

As a next step the polarization sum
\begin{equation}
\label{eq:polarization-sum}
\Pi^{\mu\nu}\equiv \sum_{\lambda=1,2}\overline{\varepsilon}^{(\lambda)\,\mu}\varepsilon^{(\lambda)\,\nu}\,,
\end{equation}
is computed where the bar means complex conjugation. To investigate the properties of the theory it is reasonable to write $\Pi^{\mu\nu}$ in a 
covariant form similar to the \textit{ansatz} of \eqref{eq:propagator-ansatz}, which was made for the propagator:
\begin{equation}
\label{eq:polarizatiom-sum-ansatz}
\Pi^{\mu\nu}=\frac{1}{N^{(6)}}\Big\{a\,\eta^{\mu\nu}+b\,k^{\mu}k^{\nu}+c\,(k^{\mu}\xi^{\nu}+\xi^{\mu}k^{\nu})+d\,\xi^{\mu}\xi^{\nu}\Big\}\Big|_{k_0=\omega}\,.
\end{equation}
Here $a$, $b$, $c$, and $d$ are unknown coefficients to be determined by comparing the \textit{ansatz} to the explicit expression
of \eqref{eq:polarization-sum} that is constructed with the polarization vectors of \eqref{eq:polarization-vectors}. Since the polarization
sum is symmetric such as the propagator this leads to ten equations that must be fulfilled.

\subsection{General results}

The propagator coefficients plus its scalar part can be computed for all ten Lorentz-violating coefficients $\kappa_{\mathrm{tr}-}^{\mu\nu}$
at once. Introducing the short-hand notation $Q^{(6)}\equiv -\kappa_{\mathrm{tr}-}^{\mu\nu}k_{\mu}k_{\nu}$
%$Q\equiv -\kappa_{\mathrm{tr}-}^{ij}k^ik^j$
one obtains:
\begin{subequations}
\label{eq:propagator-coefficients}
\begin{align}
\label{eq:scalar-propagator}
\widehat{K}&=\frac{1}{k_0^2\big(1+Q^{(6)}\big)-\mathbf{k}^2\big(1-Q^{(6)}\big)}\,, \\[2ex]
\widehat{a}&=1\,,\quad \widehat{b}=\frac{-Q^{(6)}\big[k_0^2\big(1-Q^{(6)}\big)-\mathbf{k}^2\big(1+Q^{(6)}\big)\big]}{k^4\big(1+Q^{(6)}\big)}\,, \\[2ex]
\widehat{c}&=\frac{2Q^{(6)}k_0}{k^2\big(1+Q^{(6)}\big)}\,,\quad \widehat{d}=-\frac{2Q^{(6)}}{1+Q^{(6)}}\,.
\end{align}
\end{subequations}
Note that this propagator has the same structure as the propagator that is based on the isotropic \textit{CPT}-even dimension-4
operator with the replacement $Q^{(6)}\mapsto Q^{(4)}\equiv \widetilde{\kappa}_{\mathrm{tr}}$. The minus sign in the definition of
$Q^{(6)}$ above originates from the minus sign that emerges when transforming the two additional derivatives of the dimension-6
operator to momentum space (cf. \eqref{eq:dimension-d-operator} for the general case of the dimension-$d$ operator and
\eqref{eq:dimension-six-operator} for the dimension-6 operator). I anticipate that the propagator for the full dimensional
expansion has exactly this form with $Q^{(6)}$ to be replaced by
\begin{align}
\label{eq:general-q}
Q^{(6)} \mapsto Q^{(\mathrm{full})}&\equiv \sum_{\substack{d=\text{even} \\ d\geq 4}} Q^{(d)}\equiv \sum_{\substack{d=\text{even} \\ d\geq 4}} (-1)^{d/2-2}\kappa_{\mathrm{tr}-}^{\alpha_1\dots \alpha_{(d-4)}}k_{\alpha_1}\dots k_{\alpha_{(d-4)}} \notag \\
&=\widetilde{\kappa}_{\mathrm{tr}}-\kappa_{\mathrm{tr}-}^{\alpha_1\alpha_2}k_{\alpha_1}k_{\alpha_2}+\kappa_{\mathrm{tr}-}^{\alpha_1\alpha_2\alpha_3\alpha_4}k_{\alpha_1}k_{\alpha_2}k_{\alpha_3}k_{\alpha_4}\mp \dots\,.
\end{align}
The coefficients of the polarization sum of \eqref{eq:polarizatiom-sum-ansatz} can be stated as
\begin{equation}
\label{eq:polarization-sum-coefficients}
a=-1\,,\quad b=-\frac{1}{\mathbf{k}^2}\,,\quad c=\frac{k_0}{\mathbf{k}^2}\,,\quad d=1-\frac{k_0^2}{\mathbf{k}^2}\,,
\end{equation}
where $k_0$ has to be replaced by the respective physical dispersion law. Furthermore, the normalization of the polarization vectors
is given by the following general expression:
\begin{equation}
\label{eq:normalization-general}
N^{(6)}=\frac{1}{2\omega^2}\left[\omega^2\big(1+Q^{(6)}\big)+\mathbf{k}^2\big(1-Q^{(6)}\big)\right]\,.
\end{equation}
Three remarks are in order.
First, the polarization sum of \eqref{eq:polarizatiom-sum-ansatz} together with the coefficients of \eqref{eq:polarization-sum-coefficients}
and the normalization of \eqref{eq:normalization-general} completely resembles the polarization sum of isotropic modified Maxwell theory
based on the dimension-4 operator with the replacement $\widetilde{\kappa}_{\mathrm{tr}}\mapsto -\kappa_{\mathrm{tr}-}^{\mu\nu}k_{\mu}k_{\nu}$.
Therefore, I suspect that for the isotropic {\em CPT}-even Lorentz-violating photon sector including all higher-dimensional
operators the polarization sum for each transverse mode has the same structure where $Q^{(6)}$ is replaced by the general expansion of
\eqref{eq:general-q} and $k_0$ by the corresponding modified dispersion relation.
For vanishing Lorentz violation $Q^{(6)}$ vanishes as well and $\omega=|\mathbf{k}|$. Then $N^{(6)}=1$, which shows that in the standard
case the normalization condition involving the 00-component of the energy-momentum tensor is automatically fulfilled.

Second, the terms with the coefficients $b$ and
$c$ do not play a role when $\Pi^{\mu\nu}$ is contracted with a gauge-invariant quantity. This holds due to the Ward identity,
which is still valid since the Lorentz-violating modification respects gauge invariance and no anomalies are expected to occur.
Third, for vanishing Lorentz violation we have $k_0=|\mathbf{k}|$ and, therefore, $\delta=0$. The truncated polarization sum
(meaning that all terms proportional to $k^{\mu}$ are dropped) then corresponds to the standard result \cite{PeskinSchroeder1995}
\begin{equation}
\lim_{\kappa_{\mathrm{tr}-}^{\mu\nu} \mapsto 0} \Pi^{\mu\nu}|_{\mathrm{truncated}}=-\eta^{\mu\nu}\,.
\end{equation}
Since the \textit{ans\"{a}tze} given by Eqs.~(\ref{eq:propagator-ansatz}), (\ref{eq:polarizatiom-sum-ansatz}) are sufficient to describe
the structure of the propagator and the polarization vectors, respectively, it is justified to take into account the timelike preferred
spacetime direction $\xi^{\mu}$ only. There may be more directions, which are defined by the matrix $(\kappa_{\mathrm{tr}-}^{\mu\nu})$.
However they are not needed to understand the structure of the modification.

\subsection{Temporal part}

Some of the general results presented above can be stated explicitly for the temporal part since they are not too lengthy.
For this special case there are two isotropic dispersion laws (see \eqref{eq:dispersion-law-purely-spatial}), i.e., they only depend
on $|\mathbf{k}|$. Without loss of generality, the momentum three-vector can be chosen to point along the $z$-axis. Then the transverse
polarization vectors of \eqref{eq:polarization-vectors} can be simplified to give (where one of the two possible signs is picked):
\begin{equation}
\varepsilon^{(1)\,\mu}=\frac{1}{\sqrt{N^{(6)}}}\begin{pmatrix}
0 \\
0 \\
1 \\
0 \\
\end{pmatrix}\,,\quad \varepsilon^{(2)\,\mu}=\frac{1}{\sqrt{N^{(6)}}}\begin{pmatrix}
0 \\
1 \\
0 \\
0 \\
\end{pmatrix}\,.
\end{equation}
The perturbed and the spurious mode have different normalization factors that follow from \eqref{eq:normalization-general} by
inserting the corresponding dispersion law:
\begin{subequations}
\label{eq:normalizations-temporal-case}
\begin{align}
N'^{(6)}_1&=\frac{1}{2}\left[1+\kappa_{\mathrm{tr}-}^{00}\mathbf{k}^2+\sqrt{1-\kappa_{\mathrm{tr}-}^{00}\mathbf{k}^2(6-\kappa_{\mathrm{tr}-}^{00}\mathbf{k}^2)}\right]\,, \\[2ex]
N''^{(6)}_1&=\frac{1}{2}\left[1+\kappa_{\mathrm{tr}-}^{00}\mathbf{k}^2-\sqrt{1-\kappa_{\mathrm{tr}-}^{00}\mathbf{k}^2(6-\kappa_{\mathrm{tr}-}^{00}\mathbf{k}^2)}\right]\,.
\end{align}
\end{subequations}
Note that the signs in front of the two square roots in \eqref{eq:normalizations-temporal-case} are opposite to the signs of
the square roots in the dispersion relations given by \eqref{eq:dispersion-law-purely-spatial}.

\section{Optical theorem at tree-level}
\label{sec:discussion-of-optical-theorem}

\begin{figure}[b]
\centering
\begin{equation*}\label{eq:opt-theorem}
\hspace*{-5mm}
2\,\mathrm{Im}\left(\begin{array}{c}
\includegraphics{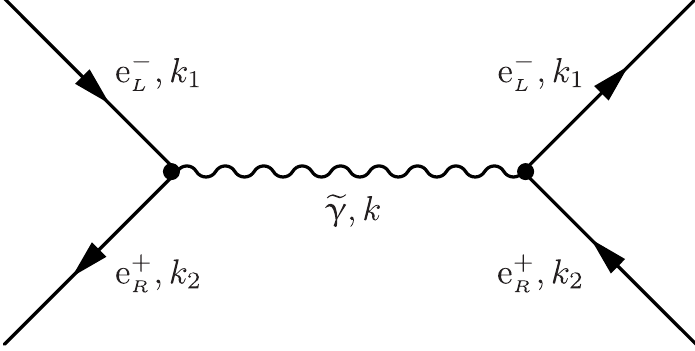}
\end{array}
\right)
\stackrel{?}{=}
\int \mathrm{d}\Pi_1 \left|\begin{array}{c}
\includegraphics{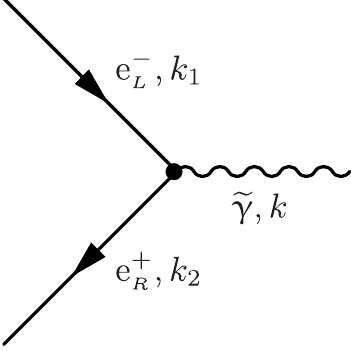}
\end{array}\right|^2\,.
\end{equation*}
\caption{Forward scattering amplitude of the process $\mathrm{e}^{-}_{L}\mathrm{e}^{+}_{R}\rightarrow \mathrm{e}^{-}_{L}\mathrm{e}^{+}_{R}$
(left-hand side) that is equal to the total cross section of $\mathrm{e}^-_L\mathrm{e}^+_R\rightarrow \widetilde{\upgamma}$ (right-hand
side) if the optical theorem is valid. Here $\widetilde{\upgamma}$ denotes a modified photon. The kinematic variables used are stated next
to the particle symbols. The infinitesimal one-particle phase space element for the process on the right-hand side of the equation is
called $\mathrm{d}\Pi_1$.}
\label{fig:optical-theorem}
\end{figure}

The occurrence of spurious photon modes in the temporal and the mixed sector of the \textit{CPT}-even modification based on the
dimension-6 operator makes us curious about the validity of the optical theorem. The latter shall be studied in the current section
where first of all the spatial case is considered. The optical theorem will be
investigated based on a particular process: the scattering of a left-handed electron and a right-handed positron at tree-level
(see \figref{fig:optical-theorem}). The calculation will be performed according to~\cite{Schreck:2011ai,Schreck:2013gma}.

As long as no problems occur in the context of the optical theorem the imaginary part of the forward scattering amplitude
$\mathcal{M}\equiv \mathcal{M}(\mathrm{e}^{-}_{L}\mathrm{e}^{+}_{R}\rightarrow \mathrm{e}^{-}_{L}\mathrm{e}^{+}_{R})$ must be related
to the production cross-section of a modified photon $\widetilde{\upgamma}$ from a left-handed electron and a right-handed positron. We
will denote the matrix
element of the latter process as $\widehat{\mathcal{M}}\equiv \mathcal{M}(\mathrm{e}_L^-\mathrm{e}_R^+\rightarrow\widetilde{\upgamma})$.
Note that it is not important, which process at tree-level is considered. In the proof no relationships will be employed that exclusively hold
for this particular process. The only property, which is assumed, is the validity of the Ward identity. This is reasonable as the axial
anomaly, which is linked to the chiral structure of quantum field theories, occurs at higher order of the electromagnetic coupling constant.

Now the forward scattering amplitude (left-hand side of \figref{fig:optical-theorem}) can be obtained with the standard Feynman rules for the
fermion sector and the modified photon propagator. It reads as follows:
\begin{align}
\label{eq:forward-scattering-amplitude-optical-theorem}
\mathcal{M}&=\int \frac{\mathrm{d}^4k}{(2\pi)^4}\,\delta^{(4)}(k_1+k_2-k)\,
e^2\;\overline{u}(k_1)\gamma^{\lambda}\frac{\mathds{1}-\gamma_5}{2}v(k_2)\;
\overline{v}(k_2)\gamma^{\nu}\frac{\mathds{1}-\gamma_5}{2}u(k_1) \notag \displaybreak[0]\\
&\phantom{{}={}\int \frac{\mathrm{d}^4k}{(2\pi)^4}}\,\times\frac{1}{\widehat{K}^{-1}+\mathrm{i}\epsilon}\;
\big[\widehat{a}\,\eta_{\nu\lambda}
+\widehat{b}\,k_{\nu}k_{\lambda}
+\widehat{c}\,(k_{\nu}\xi_{\lambda}+\xi_{\nu}k_{\lambda})
+\widehat{d}\,\xi_{\nu}\xi_{\lambda}
\big]\,.
\end{align}
Here $e$ is the elementary charge, $u$, $v$, $\overline{u}$, and $\overline{v}$ are standard Dirac spinors,
$\gamma_5=\mathrm{i}\gamma^0\gamma^1\gamma^2\gamma^3$ with the standard Dirac matrices $\gamma^{\mu}$
(for $\mu\in \{0,1,2,3\}$), and $\mathds{1}$ is the unit matrix in spinor space. The kinematical variables used are
shown in \figref{fig:optical-theorem}. The four-dimensional $\delta$-function ensures total four-momentum conservation.
The photon propagator with the propagator coefficients is taken from \eqref{eq:propagator-coefficients}. The physical poles that
appear in the scalar propagator part are treated with the ordinary $\mathrm{i}\epsilon$-procedure. This means that the positive pole is
shifted to the lower complex half-plane (meaning that an integration contour runs above the pole) and the negative one is shifted to the
upper complex half-plane (where a contour runs below this pole).

\subsection{Spatial part}
\label{sec:optical-theorem-purely-spatial}

For the spatial part of the modified theory considered the procedure used in \cite{Schreck:2011ai,Schreck:2013gma} does not
fundamentally change. The denominator of the scalar propagator part is factorized with respect to the propagator poles. Terms of
quadratic and higher order in the infinitesimal parameter $\epsilon$ are not taken into account. The photon propagator has two
physical poles, where
\begin{equation}
\label{eq:dispersion-relation-purely-spatial}
\omega=\sqrt{\frac{1-Q_3^{(6)}}{1+Q_3^{(6)}}}\,|\mathbf{k}|\,,\quad Q^{(6)}_3\equiv -\kappa^{ij}_{\mathrm{tr}-}k^ik^j\,,
\end{equation}
is the positive one and $-\omega<0$ corresponds to its negative counterpart. The scalar part $\widehat{K}$ of the propagator is then
written in the following form:
\begin{align}
\frac{1}{\widehat{K}^{-1}+\mathrm{i}\epsilon}&=\frac{1}{k_0^2\big(1+Q^{(6)}_3\big)-\mathbf{k}^2\big(1-Q^{(6)}_3\big)+\mathrm{i}\epsilon} \notag \\
&=\frac{1}{\big(1+Q^{(6)}_3\big)(k^0-\omega+\mathrm{i}\epsilon)(k^0+\omega-\mathrm{i}\epsilon)}\,.
\end{align}
Due to the $\mathrm{i}\epsilon$-prescription the relation
\begin{equation}
\label{eq:propagator-relations-principal-value}
\frac{1}{k^0-\omega+\mathrm{i}\epsilon}=\mathcal{P}\frac{1}{k^0-\omega}-\mathrm{i}\pi\delta(k^0-\omega)\,,
\end{equation}
with the principal value $\mathcal{P}$ holds for the physical pole. The first part of \eqref{eq:propagator-relations-principal-value} is
purely real. The second part
is imaginary and due to the $\delta$-function it forces the zeroth four-momentum component to be equal to the respective physical photon
frequency. The negative pole does not contribute to the imaginary part because of total four-momentum conservation. Furthermore
\begin{equation}
\left.\frac{1}{\big(1+Q^{(6)}_3\big)(k^0+\omega+\mathrm{i}\epsilon)}\right|_{k^0=\omega}=\frac{1}{2\omega N^{(6)}_3}\,,\quad N^{(6)}_3\equiv 1-\kappa_{\mathrm{tr}-}^{ij}k^ik^j\,,
\end{equation}
where $N^{(6)}_3$ follows from \eqref{eq:normalization-general} by inserting the dispersion relation of \eqref{eq:dispersion-relation-purely-spatial}.
Using these results, the $k^0$-integration in \eqref{eq:forward-scattering-amplitude-optical-theorem} can be done. Since the interest lies
in the imaginary part, terms involving the principal value are not considered and $k^0$ is replaced by the photon frequency~$\omega$:
\begin{subequations}
\begin{align}\label{eq:optical-theorem-mode-1}
2\,\mathrm{Im}(\mathcal{M})&=\int
\frac{\mathrm{d}^3k}{(2\pi)^{3}\,2\omega}\,\delta^{(4)}(k_1+k_2-k) \notag \\
&\hspace{1.5cm}\times e^2\;\overline{u}(k_1)\gamma^{\nu}\frac{\mathds{1}-\gamma_5}{2}v(k_2)\;
\overline{v}(k_2)\gamma^{\mu}\frac{\mathds{1}
-\gamma_5}{2}u(k_1)\,\frac{1}{N^{(6)}_3}\;(-\eta_{\mu\nu}+d\,\xi^{\mu}\xi^{\nu})
\notag \displaybreak[0]\\
&=\int \frac{\mathrm{d}^3k}{(2\pi)^{3}\,2\omega}\,\delta^{(4)}(k_1+k_2-k)\,
(\widehat{\mathcal{M}}^{\,\dagger})^{\nu}(\widehat{\mathcal{M}})^{\mu}\Pi_{\mu\nu}
\notag \displaybreak[0]\\[1mm]
&=\int
\frac{\mathrm{d}^3k}{(2\pi)^{3}\,2\omega}\,\delta^{(4)}(k_1+k_2-k)|\widehat{\mathcal{M}}|^2\,,
\end{align}
with
\begin{equation}
\widehat{\mathcal{M}}\equiv \sum_{\lambda=1,2} \varepsilon^{(\lambda)}_{\mu}(k)(\widehat{\mathcal{M}})^{\mu}(k)\,.
\end{equation}
\end{subequations}
Terms that involve at least one four-momentum in the tensor structure of the propagator can be dropped, if the Ward
identity is taken into account. Hence, the optical theorem at tree-level is valid for the spatial sector as expected.

\subsection{Temporal part}

The temporal part is characterized by two distinct dispersion relations according to \eqref{eq:dispersion-law-purely-spatial}.
To make the following calculations more transparent they will be written as follows:
\begin{equation}
\label{eq:dispersion-relations-purely-temporal}
\omega_{1,2}(\mathbf{k})=\frac{1}{\sqrt{2\kappa_{\mathrm{tr}-}^{00}}}\sqrt{a\mp \sqrt{b}}\,,\quad a=1-\kappa_{\mathrm{tr}-}^{00}\mathbf{k}^2\,,\quad b=1-\kappa_{\mathrm{tr}-}^{00}\mathbf{k}^2(6-\kappa_{\mathrm{tr}-}^{00}\mathbf{k}^2)\,.
\end{equation}
The first of these is a perturbation of the standard dispersion law whereas this is not the case for the second. Therefore the second
can be considered as spurious for momenta that are much smaller than the Planck scale. Nevertheless there is no reason why it formally
should not be taken into account in the optical theorem. Although it is considered as spurious it is, indeed, a transverse dispersion law
(see the discussion in \secref{sec:dispersion-relations-purely-temporal}). When $\omega_2$ is not discarded, the structure of the temporal
sector is reminiscent of the structure of a birefringent theory. In \cite{Schreck:2013gma} a birefringent sector of modified Maxwell theory
was considered that is based on the dimension-4 operator. If both $\omega_1$ and $\omega_2$ are assumed to contribute to the imaginary part
of the forward scattering amplitude the calculation can be performed analogously to how this was done in the latter reference. The only
difference is that each transverse dispersion relation is linked to a separate polarization sum according to \eqref{eq:polarization-sum}
and not only to one of the two contributions of the sum. This has to do with the twofold degeneracy of each dispersion law (cf. the last
paragraph of \secref{sec:dispersion-relations-classical-theory}).

The scalar part of the modified photon propagator then has four different poles. Two of them are given by \eqref{eq:dispersion-relations-purely-temporal}
and the other two by their respective negative counterparts. The denominator of the scalar part is factorized with respect to these poles and
the $\mathrm{i}\epsilon$-prescription is applied again:
\begin{equation}
\frac{1}{\widehat{K}^{-1}+\mathrm{i}\epsilon}=-\frac{1}{\kappa_{\mathrm{tr}-}^{00}(k^0-\omega_1+\mathrm{i}\epsilon)(k^0+\omega_1-\mathrm{i}\epsilon)(k^0-\omega_2+\mathrm{i}\epsilon)(k^0+\omega_2-\mathrm{i}\epsilon)}\,.
\end{equation}
Then \eqref{eq:propagator-relations-principal-value} can be used for each of the positive poles. The negative ones do not play a role due
to four-momentum conservation. This results in the following contributions to the imaginary part, where a factor of $-\mathrm{i}\pi$ is omitted:
\begin{subequations}
\begin{align}
\label{eq:optical-theorem-temporal-lhs-1}
\left.-\frac{1}{\kappa_{\mathrm{tr}-}^{00}(k^0+\omega_1)(k^0-\omega_2)(k^0+\omega_2)}\right|_{k^0=\omega_1}&=-\frac{1}{2\kappa_{\mathrm{tr}-}^{00}\omega_1(\omega_1^2-\omega_2^2)}=\frac{\sqrt{\kappa_{\mathrm{tr}-}^{00}}}{\sqrt{2}\sqrt{a-\sqrt{b}}\sqrt{b}}\,, \\[2ex]
\label{eq:optical-theorem-temporal-lhs-2}
\left.-\frac{1}{\kappa_{\mathrm{tr}-}^{00}(k^0-\omega_1)(k^0+\omega_1)(k^0+\omega_2)}\right|_{k^0=\omega_2}&=-\frac{1}{2\kappa_{\mathrm{tr}-}^{00}\omega_2(\omega_2^2-\omega_1^2)} \notag \\
&=-\frac{\sqrt{\kappa_{\mathrm{tr}-}^{00}}}{\sqrt{2}\sqrt{a+\sqrt{b}}\sqrt{b}}\,.
\end{align}
\end{subequations}
According to the previous section and the discussion in \cite{Schreck:2013gma}, for the optical theorem to be valid these expressions have
to correspond to the following results where $N'^{(6)}_1$ and $N''^{(6)}_1$ are the normalizations of the temporal case taken from
\eqref{eq:normalizations-temporal-case}:
\begin{subequations}
\begin{align}
\label{eq:optical-theorem-temporal-rhs-1}
\frac{1}{4\omega_1N'^{(6)}_1}&=\frac{\sqrt{\kappa_{\mathrm{tr}-}^{00}}}{\sqrt{2}\sqrt{a-\sqrt{b}}}\frac{2}{2-a+\sqrt{b}+2\left[1+2/(a-\sqrt{b})\right]\kappa_{\mathrm{tr}-}^{00}\mathbf{k}^2}\,, \\[2ex]
\label{eq:optical-theorem-temporal-rhs-2}
\frac{1}{4\omega_2N''^{(6)}_1}&=\frac{\sqrt{\kappa_{\mathrm{tr}-}^{00}}}{\sqrt{2}\sqrt{a+\sqrt{b}}}\frac{2}{2-a-\sqrt{b}+2\left[1+2/(a+\sqrt{b})\right]\kappa_{\mathrm{tr}-}^{00}\mathbf{k}^2}\,.
\end{align}
\end{subequations}
However, Eqs.~(\ref{eq:optical-theorem-temporal-lhs-1}), (\ref{eq:optical-theorem-temporal-rhs-1}) and Eqs.~(\ref{eq:optical-theorem-temporal-lhs-2}),
(\ref{eq:optical-theorem-temporal-rhs-2}) can evidently not be equal to each other. Therefore the optical theorem at tree-level appears to be violated
for the temporal sector of the dimension-6 operator. The reason is the occurrence of a spurious dispersion law, which renders the
structure of the theory birefringent.
%This is in contract to the initially claimed property of being an isotropic modification.

\subsection{Mixed case}

Also the mixed case has two distinct dispersion relations that are given by Eqs.~(\ref{eq:dispersion-laws-transverse-1-total}),
(\ref{eq:dispersion-laws-transverse-2-total}) where only one of them is a perturbation of the standard dispersion law. Contrary to the temporal sector, they change their roles dependent on the choice of Lorentz-violating coefficients and momentum components (see the
discussion in \secref{subsec:mixed-part}). Without loss of generality in what follows we assume that $\omega_1$ is the perturbation
and $\omega_2>0$ is spurious.

The scalar propagator part has three poles: $\omega_1$, $\omega_2$, and a third negative pole $\omega_3$. The latter is neither a negative
counterpart of $\omega_1$ nor of $\omega_2$. The denominator of the scalar propagator is then factorized as follows:
\begin{align}
\frac{1}{\widehat{K}^{-1}+\mathrm{i}\epsilon}&=\frac{1}{a^{-1}k_0^3+k_0^2+a^{-1}\mathbf{k}^2k_0-\mathbf{k}^2+\mathrm{i}\epsilon} \notag \\
&=\frac{1}{a^{-1}(k^0-\omega_1+\mathrm{i}\epsilon)(k^0-\omega_2+\mathrm{i}\epsilon)(k^0+\omega_3-\mathrm{i}\epsilon)}\,.
\end{align}
Here $a^{-1}=2\kappa_{\mathrm{tr}-}^{0i}k^i$ is the inverse of $a$ defined in \eqref{eq:transformation-cardano}.
According to \eqref{eq:propagator-relations-principal-value} the poles $k^0=\omega_1$ and $k^0=\omega_2$ deliver the following contributions
to the imaginary part where a factor $-\mathrm{i}\pi$ has again been dropped:
\begin{subequations}
\begin{align}
\label{eq:optical-theorem-mixed-lhs-1}
\left.\frac{1}{a^{-1}(k^0-\omega_2)(k^0+\omega_3)}\right|_{k^0=\omega_1}&=\frac{1}{a^{-1}(\omega_1-\omega_2)(\omega_1+\omega_3)}\,, \\[2ex]
\label{eq:optical-theorem-mixed-lhs-2}
\left.\frac{1}{a^{-1}(k^0-\omega_1)(k^0+\omega_3)}\right|_{k^0=\omega_2}&=\frac{1}{a^{-1}(\omega_2-\omega_1)(\omega_2+\omega_3)}\,.
\end{align}
\end{subequations}
If the optical theorem is valid, each of these expressions must correspond to the respective following contribution:
\begin{subequations}
\begin{align}
\label{eq:optical-theorem-mixed-rhs-1}
\frac{1}{4\omega_1N'^{(6)}_2}&=\frac{\omega_1}{2\left[\omega_1^2(1+\omega_1/a)+\mathbf{k}^2(1-\omega_1/a)\right]}\,, \\[2ex]
\label{eq:optical-theorem-mixed-rhs-2}
\frac{1}{4\omega_2N''^{(6)}_2}&=\frac{\omega_2}{2\left[\omega_2^2(1+\omega_2/a)+\mathbf{k}^2(1-\omega_2/a)\right]}\,.
\end{align}
\end{subequations}
There are again two different normalization factors such as in the temporal case that are denoted as $N'^{(6)}_2$
and $N''^{(6)}_2$. They result from \eqref{eq:normalization-general} by inserting the appropriate dispersion law $\omega_1$
or $\omega_2$.
Since Eqs.~(\ref{eq:optical-theorem-mixed-lhs-1}), (\ref{eq:optical-theorem-mixed-lhs-2}) depend on $\omega_3$, which does not
appear in Eqs.~(\ref{eq:optical-theorem-mixed-rhs-1}), (\ref{eq:optical-theorem-mixed-rhs-2}), these results cannot correspond to
each other. This can also be explicitly shown by inserting the modified dispersion relations directly. Therefore the optical
theorem at tree-level seems to be invalid for the mixed sector as well.

\subsection{Interpretation}

In the last section it was demonstrated that the optical theorem at tree-level appears to be violated for the temporal and the
mixed case of the Lorentz-violating modification based on the {\em CPT}-even dimension-6 operator in the photon sector. However a
violation of the optical theorem does not necessarily indicate that unitarity is violated as well.\footnote{For example, see
\cite{Kupczynski:2013doa} and references therein for an argumentation why the optical theorem could, indeed, be violated for
processes mediated by the strong interaction, whereas unitarity of the S-matrix is still granted.}

Since the optical theorem is perfectly valid for the spatial case its violation for the other two sectors must be linked
to the additional time derivatives that appear in the higher-dimensional operator. These time derivatives may spoil the Hamiltonian
of the system, which renders the time evolution of the states unconventional. This was pointed out to happen in the fermion sector
of the SME \cite{Colladay:2001wk}. The observation made here is supposed to be the analogue in the nonminimal photon sector. In the
latter reference it was shown that the issues arising can be solved at first order Lorentz violation by a field redefinition.

Here we will go an alternative path. The field is not redefined but the additional time derivatives (or the $k_0$-components in
momentum space) can be eliminated from the field equations at first order Lorentz violation by taking into account that
\begin{equation}
k^2=k_0^2-\mathbf{k}^2=\mathcal{O}(\kappa_{\mathrm{tr}-}^{\mu\nu}k_{\mu}k_{\nu})\,,
\end{equation}
where $\kappa_{\mathrm{tr}-}^{\mu\nu}k_{\mu}k_{\nu}$ contains a particular subset of Lorentz-violating coefficients. Each additional
$k_0$, which appears in the dimension-6 operator, is multiplied with a Lorentz-violating coefficient. Hence, replacing these $k_0$
by $|\mathbf{k}|$ leads to an error at {\em second} order Lorentz violation whereas all expressions remain valid at first order.

\subsubsection{Temporal case}

For this case the replacement $k_0^2\kappa_{\mathrm{tr}-}^{00}\mapsto \mathbf{k}^2\kappa_{\mathrm{tr}-}^{00}$ is performed in the
matrix $M$ of \eqref{eq:field-equations-modified-maxwell-theory}.
%\begin{equation}
%k_0^3|\mathbf{k}|\kappa_{\mathrm{tr}-}^{00}\mapsto k_0|\mathbf{k}|^3\kappa_{\mathrm{tr}-}^{00}\,,\quad
%k_0^4\kappa_{\mathrm{tr}-}^{00}\mapsto k_0^2\mathbf{k}^2\kappa_{\mathrm{tr}-}^{00}\,,\quad k_0^2\mathbf{k}^2\kappa_{\mathrm{tr}-}^{00}\mapsto \mathbf{k}^4\kappa_{\mathrm{tr}-}^{00}\,.
%\end{equation}
Notice that no spurious dispersion law appears any more after this replacement has been made. One obtains a
single isotropic modified dispersion relation that is given by:
\begin{equation}
\omega(\mathbf{k})=\sqrt{\frac{1-Q^{(6)}_1}{1+Q^{(6)}_1}}\,|\mathbf{k}|\,,\quad Q^{(6)}_1\equiv -\kappa_{\mathrm{tr}-}^{00}\mathbf{k}^2\,.
\end{equation}
With the modified dispersion law in this form the optical theorem can be shown to be valid in a completely analogous way as this was
done for the spatial case in \secref{sec:optical-theorem-purely-spatial}. The only thing to do is to replace $Q^{(6)}_3$ by
$Q^{(6)}_1$ defined above. The caveat is that this calculation is only applicable at first order Lorentz violation due to the performed
replacement.

\subsubsection{Mixed case}

For the mixed case the replacement $k_0\kappa_{\mathrm{tr}-}^{0i}\mapsto |\mathbf{k}|\kappa_{\mathrm{tr}-}^{0i}$ is done in the matrix $M$ of
\eqref{eq:field-equations-modified-maxwell-theory}. Also here the spurious dispersion law disappears and one ends up with a single
modified dispersion relation:
\begin{equation}
\omega(\mathbf{k})=\sqrt{\frac{1-Q^{(6)}_2}{1+Q^{(6)}_2}}\,|\mathbf{k}|\,,\quad Q^{(6)}_2\equiv 2\kappa_{\mathrm{tr}-}^{0i}|\mathbf{k}|k^i\,.
\end{equation}
In this case the optical theorem can be shown to be valid with the transformed dispersion law following the lines of
\secref{sec:optical-theorem-purely-spatial} with $Q^{(6)}_3$ replaced by $Q^{(6)}_2$. Therefore the validity optical theorem at tree-level
has been demonstrated for this sector at first order Lorentz violation as well.

\section{Gauge field commutator}
\label{sec:gauge-field-commutator}

In the current section the interest lies in the commutator $K^{\mu\nu}(y,z)\equiv [A^{\mu}(y),A^{\nu}(z)]$ of two gauge fields where
both are evaluated at different spacetime points $y$ and $z$. This
commutator is closely linked to the causal structure of the theory. If it vanishes for two distinct spacetime points
the physical fields can be measured exactly at the respective points. A nonvanishing commutator indicates that measurements of
physical observables at the corresponding spacetime points can influence each other. In this case they cannot be simultaneously
measured with arbitrary precision because of the uncertainty relation.

Due to translational invariance of the theory a coordinate transformation $w\mapsto w'=w-z$ can be performed such that it suffices
to consider $K^{\mu\nu}(x)\equiv K^{\mu\nu}(x,0)=[A^{\mu}(x),A^{\nu}(0)]$ with $x\equiv y-z$ instead. The goal is to gain some
understanding in the commutator without too much calculational effort. For this reason the general case with ten Lorentz-violating
coefficients is reduced to the spatial case with the equal coefficients
$\kappa_{\mathrm{tr}-}^{11}=\kappa_{\mathrm{tr}-}^{22}=\kappa^{33}_{\mathrm{tr}-}\equiv \overline{\kappa}_{\mathrm{tr}}$
and all others set to zero. With
\begin{equation}
Q^{(6)}_3=-(\kappa_{\mathrm{tr}-}^{11}k_1^2+\kappa_{\mathrm{tr}-}^{22}k_2^2+\kappa_{\mathrm{tr}-}^{33}k_3^2)=-\overline{\kappa}_{\mathrm{tr}}\mathbf{k}^2\,,
\end{equation}
the following isotropic dispersion relation is obtained:
\begin{equation}
\label{eq:modified-dispersion-law-isotropic}
\omega=\sqrt{\frac{1+\overline{\kappa}_{\mathrm{tr}}\mathbf{k}^2}{1-\overline{\kappa}_{\mathrm{tr}}\mathbf{k}^2}}\,|\mathbf{k}|\,.
\end{equation}
The commutator can be written in the form $K^{\mu\nu}(x)=\mathrm{i}\theta^{\mu\nu}(x)D(x)$ where $\theta^{\mu\nu}$ respects its tensor
structure and $D(x)$ is the scalar commutator function. In App.~\ref{sec:gauge-field-commutator-calculation} it is demonstrated that
$\theta^{\mu\nu}$ is related to the polarization sum and $D(x)$ involves the scalar propagator function $\widehat{K}$ of
\eqref{eq:scalar-propagator}. The scalar commutator function can then be computed as a contour integral in the complex $k^0$-plane
followed by an integration over the three-dimensional momentum space. The contour $C$ has to be chosen such that all poles of the integrand
are encircled in counterclockwise direction. The $k^0$-integral can then be performed with the residue theorem:
\begin{align}
D(x)&=\oint_C \frac{\mathrm{d}k^4}{(2\pi)^4}\,\widehat{K}\exp(-\mathrm{i}kx)=\int_C \frac{\mathrm{d}k^0}{2\pi} \int\frac{\mathrm{d}k^3}{(2\pi)^3}\,\frac{\exp(-\mathrm{i}k^0x^0+\mathrm{i}\mathbf{k}\cdot \mathbf{x})}{(1-\overline{\kappa}_{\mathrm{tr}}\mathbf{k}^2)(k^0-\omega)(k^0+\omega)} \notag \\
&=\int \frac{\mathrm{d}^3k}{(2\pi)^3}\,\frac{1}{1-\overline{\kappa}_{\mathrm{tr}}\mathbf{k}^2}\frac{\mathrm{i}}{2\omega}\left[\exp(-\mathrm{i}\omega x^0)-\exp(\mathrm{i}\omega x^0)\right]\exp(\mathrm{i}\mathbf{k}\cdot \mathbf{x})\,.
\end{align}
Note that the sign convention in the complex exponential function differs from the convention used in \cite{Klinkhamer:2010zs,Schreck:2011ai}
and it is in accordance to \cite{Schreck:2013gma}. At this point it is reasonable to introduce spherical coordinates due to the isotropy of
the case considered. This leads to:
\begin{align}
\label{eq:scalar-commutator-function-modified}
D(x)&=\frac{1}{(2\pi)^2} \int_0^{\pi} \mathrm{d}\vartheta\,\sin\vartheta\int_0^{\infty} \mathrm{d}|\mathbf{k}|\,\frac{1}{1-\overline{\kappa}_{\mathrm{tr}}|\mathbf{k}|^2}\frac{|\mathbf{k}|^2}{\omega}\sin(\omega x^0)\exp(\mathrm{i}|\mathbf{k}||\mathbf{x}|\cos\vartheta) \notag \\
%&=\frac{1}{(2\pi)^2|\mathbf{x}|} \int_0^{\infty} \mathrm{d}|\mathbf{k}|\,\frac{1}{1-\overline{\kappa}_{\mathrm{tr}}|\mathbf{k}|^2}\frac{|\mathbf{k}|}{\omega}\sin(\omega x^0)\frac{1}{\mathrm{i}}\left[\exp(\mathrm{i}|\mathbf{k}||\mathbf{x}|)-\exp(-\mathrm{i}|\mathbf{k}||\mathbf{x}|)\right] \notag \\
&=\frac{1}{2\pi^2|\mathbf{x}|} \int_0^{\infty} \mathrm{d}|\mathbf{k}|\,\frac{1}{1-\overline{\kappa}_{\mathrm{tr}}|\mathbf{k}|^2}\frac{|\mathbf{k}|}{\omega}\sin(\omega x^0)\sin(|\mathbf{k}||\mathbf{x}|) \notag \\
&=\frac{1}{2\pi^2|\mathbf{x}|} \int_0^{\infty} \mathrm{d}|\mathbf{k}|\,\frac{1}{\sqrt{1-\overline{\kappa}_{\mathrm{tr}}^2|\mathbf{k}|^4}}\sin\left(\sqrt{\frac{1+\overline{\kappa}_{\mathrm{tr}}|\mathbf{k}|^2}{1-\overline{\kappa}_{\mathrm{tr}}|\mathbf{k}|^2}}|\mathbf{k}|x^0\right)\sin(|\mathbf{k}||\mathbf{x}|)\,.
\end{align}
For $\overline{\kappa}_{\mathrm{tr}}=0$ the integral can be computed to give the standard result \cite{JordanPauli1928,Heitler1954}
\begin{equation}
\label{eq:scalar-commutator-function-standard}
\Delta(x)\equiv \lim_{\overline{\kappa}_{\mathrm{tr}}\mapsto 0} D(x)=\frac{1}{2\pi}\mathrm{sgn}(x^0)\delta\left[(x^0)^2-|\mathbf{x}|^2\right]\,,\quad \mathrm{sgn}(x)=\left\{\begin{array}{rcl}
-1 & \text{for} & x<0\,, \\
0 & \text{for} & x=0\,, \\
1 & \text{for} & x>0\,. \\
\end{array}
\right.
\end{equation}
The computation of the integral in \eqref{eq:scalar-commutator-function-modified} for nonvanishing $\overline{\kappa}_{\mathrm{tr}}$ is
prohibitively difficult. However note that an
evaluation of this integral is not necessarily reasonable. This is because the theory considered is effective and assumed to break
down when $|\mathbf{k}|$ lies in the vicinity of the Planck scale. Thus the form of the integrand cannot be assumed to be valid for
arbitrarily large integration momentum $|\mathbf{k}|$. This is directly evident from the properties of the dispersion law and the
integrand. For $|\mathbf{k}|>1/\sqrt{\overline{\kappa}_{\mathrm{tr}}}$ the square roots have branch cuts and the dispersion relation is
complex-valued. Photons can then not propagate any more and the physical meaning of $D(x)$ is lost. Therefore the integration should
be cut-off at some value $|\mathbf{k}|=\Lambda\ll 1/\sqrt{\overline{\kappa}_{\mathrm{tr}}}$.

Nevertheless even if the integral is not evaluated some of its properties can be deduced. The full result for the commutator given by
\eqref{eq:commutator-vector-potentials} is a solution of the modified free field equations (\ref{eq:field-equations-modified-maxwell-theory})
of the photon sector. Furthermore it holds that $D(x)$ vanishes for $x^0=0$ and that $D(x)=D(-x)$. These properties are valid for
the standard result $\Delta(x)$ as well. However, many of the remaining properties that hold for the standard theory are spoiled
for the effective theory considered here due to the singularity of the integrand for $|\mathbf{k}|=1/\sqrt{\overline{\kappa}_{\mathrm{tr}}}$.
The latter changes the small-distance behavior of the theory.
If the integration is done to the cut-off $\Lambda\ll 1/\sqrt{\overline{\kappa}_{\mathrm{tr}}}$ then $\overline{\kappa}_{\mathrm{tr}}|\mathbf{k}|^2\ll 1$
can be exploited. Under this assumption the integrand is expanded where terms of second and higher order in $\overline{\kappa}_{\mathrm{tr}}$
are discarded:
\begin{subequations}
\label{eq:scalar-commutator-function-modified-expanded}
\begin{align}
D(x)&=\frac{1}{2\pi^2|\mathbf{x}|} \int_0^{\Lambda} \mathrm{d}|\mathbf{k}|\,\left[\sin(|\mathbf{k}|x^0)\sin(|\mathbf{k}||\mathbf{x}|)+\overline{\kappa}_{\mathrm{tr}}|\mathbf{k}|^3x^0\cos(|\mathbf{k}|x^0)\sin(|\mathbf{k}||\mathbf{x}|)\right]+\mathcal{O}(\overline{\kappa}_{\mathrm{tr}}^2) \notag \\
&=\frac{1}{2\pi^2|\mathbf{x}|}\left(f(\Lambda)-\overline{\kappa}_{\mathrm{tr}}x^0\frac{\mathrm{d}^3f(\Lambda)}{\mathrm{d}(x^0)^3}+\mathcal{O}(\overline{\kappa}_{\mathrm{tr}}^2)\right)\,, \\[2ex]
f(|\mathbf{k}|)&=\frac{x^0\sin[|\mathbf{k}|(x^0-|\mathbf{x}|)]}{(x^0)^2-|\mathbf{x}|^2}-\frac{\sin(|\mathbf{k}|x^0)\cos(|\mathbf{k}||\mathbf{x}|)}{x^0+|\mathbf{x}|}\,.
\end{align}
\end{subequations}
The standard scalar commutator function $\Delta(x)$, which is given by \eqref{eq:scalar-commutator-function-standard}, is a
distribution. Therefore, $D(x)$ is interpreted as a distribution as well. Note that the trigonometric functions occurring in
\eqref{eq:scalar-commutator-function-modified-expanded} oscillate very rapidly since they are evaluated at the cut-off $\Lambda$.
If $D(x)$ is multiplied with a sufficiently smooth
function $g(x)$ and the product is integrated over $x$, the main contribution to the result comes from the regions
$x^0=|\mathbf{x}|$ and $x^0=-|\mathbf{x}|$. These define the standard nullcone in configuration space. This is exactly the
behavior that one would expect from a Lorentz-violating theory where Lorentz violation is assumed to be a small perturbation.

Contrary to the power-counting renormalizable cases of modified Maxwell theory previously considered in
\cite{Casana-etal2009,Casana-etal2010,Klinkhamer:2010zs,Schreck:2011ai,Schreck:2013gma} the modified
nullcones in configuration space cannot be determined from the scalar commutator function. The reason is that the theory of
nonrenormalizable dimension, which is considered in this paper, is expected to be valid only for distances much larger than 
$\sqrt{\overline{\kappa}_{\mathrm{tr}}}$. Such distances correspond to small momenta and deviations from the standard dispersion 
relation are small according to \eqref{eq:modified-dispersion-law-isotropic}. That is why only the standard nullcone structure is 
encoded in \eqref{eq:scalar-commutator-function-modified-expanded}.

\section{Conclusions}
\label{sec:conclusions}

In the current article the Lorentz-violating extended QED based on the dimension-6 operator of the {\em CPT}-even modified Maxwell term was
investigated. This extension can be considered as an effective quantum field theory of nonrenormalizable dimension that is predictive as 
long as the photon momentum is much smaller than the Planck scale.
The originally ten Lorentz-violating coefficients $\kappa_{\mathrm{tr}-}^{\mu\nu}$ were grouped into three different sectors:
the temporal one with the only coefficient $\kappa_{\mathrm{tr}-}^{00}$, the mixed case with the three coefficients
$\kappa_{\mathrm{tr}-}^{0i}$ for $i=1$, 2, 3 plus the spatial one with the remaining six coefficients.

Apart from the modified photon dispersion relation being a perturbation of the standard dispersion law, both the temporal and the mixed
sector are characterized by additional spurious dispersion relations. These are not perturbations but for small Lorentz violation they
contain terms proportional to some negative power of the Lorentz-violating coefficient. If they are taken into account the modifications
show characteristics of birefringent photon theories.

Having obtained the modified dispersion laws, the goal was to test the validity of the optical theorem due to the peculiar properties
of the Lorentz-violating modifications. The upshot is that the optical theorem at tree-level was found to hold for the spatial case
whereas it seemed to be violated for the remaining two cases. The violation was traced back to the additional time derivatives that
occur in the dimension-6 operator. After removing these time derivatives by a replacement rule that is valid at first order Lorentz
violation, it was demonstrated that both the spurious modes are removed and the validity of the optical theorem at tree-level is
restored.

Furthermore for a particular case of Lorentz-violating coefficients it was shown that the commutator of two gauge fields
evaluated at different spacetime points can be reduced to a one-dimensional integral. The integrand becomes singular for
momenta lying in the order of magnitude of the inverse square root of the Lorentz-violating coefficient, which may probably be
associated with the Planck scale. Hence the nonrenormalizable theory breaks down for such momentum scales and, therefore, the
short-distance behavior of the modified nullcone in configuration space cannot be determined. However, for distances much larger
than the Planck length the nullcone can be considered as standard.

\section{Acknowledgments}

It is a pleasure to thank V.~A.~Kosteleck\'{y} for very helpful comments and suggestions on the first version of the manuscript. This
work was performed with financial support from the \textit{Deutsche Akademie der Naturforscher Leopoldina} within Grant No. LPDS
2012-17.

%\newpage
\begin{appendix}
\numberwithin{equation}{section}

\section{Normalization of polarization vectors}
\label{sec:normalization-polarization-vectors}

The normalization of the photon polarization vectors does not follow from the field equations but from the condition
\begin{equation}
\label{eq:normalization-condition}
\langle \mathbf{k},\sigma|:P^0:|\mathbf{k},\sigma\rangle=\langle \mathbf{k},\sigma|\int \mathrm{d}^3x\,:T^{00}:|\mathbf{k},\sigma\rangle \overset{!}{=} \omega(\mathbf{k})\,.
\end{equation}
Here $|\mathbf{k},\sigma\rangle$ is a state describing a single photon with three-momentum $\mathbf{k}$ and transverse
polarization~$\sigma$. The integration is performed over three-dimensional configuration space.
Furthermore, $:T^{00}:$ is the normal-ordered 00-component of the energy-momentum tensor, which in modified Maxwell theory
reads as follows \cite{Colladay:1998fq}:
\begin{equation}
\label{eq:energy-momentum-tensor}
T^{00}=\frac{1}{2}(\mathbf{E}^2+\mathbf{B}^2)-(k_F)^{0j0k}E^jE^k+\frac{1}{4}(k_F)^{jklm}\varepsilon^{jkp}\varepsilon^{lmq}B^pB^q\,,
\end{equation}
where $\mathbf{E}=(E^1,E^2,E^3)$ and $\mathbf{B}=(B^1,B^2,B^3)$ are the electric and magnetic field strength vectors, respectively.
The symbol $\varepsilon^{ijk}$ denotes the totally antisymmetric Levi-Civita tensor. The electric and magnetic fields can be
obtained from the vector potential $A^{\mu}$. We write the latter as a Fourier decomposition with annihilation operators $a(\mathbf{k})$
and creation operators $a^{\dagger}(\mathbf{k})$:
\begin{subequations}
\label{eq:vector-potential}
\begin{align}
A^{\mu}(x)&=\sum_{r=1,2}\int \widetilde{\mathrm{d}k}\,\left[a^{(r)}(k)\varepsilon^{(r)\,\mu}(\mathbf{k})\exp(-\mathrm{i}kx)+a^{(r)\,\dagger}(k)\overline{\varepsilon}^{(r)\,\mu}(\mathbf{k})\exp(\mathrm{i}kx)\right]\,, \\[2ex]
\widetilde{\mathrm{d}k}&=\frac{\mathrm{d}^3k}{(2\pi)^32\omega(\mathbf{k})}\,,
\end{align}
\end{subequations}
with $k^0=\omega(\mathbf{k})$. The bar denotes complex conjugation and the summation runs over the physical polarizations $r=1$, 2.
The $\mathbf{E}$- and $\mathbf{B}$-fields follow from the vector potential in the usual manner:
\begin{subequations}
\begin{align}
\mathbf{E}&=-\frac{\partial\mathbf{A}}{\partial t}-\boldsymbol{\nabla}A^0 \notag \\
&=\sum_{r=1,2} \int \widetilde{\mathrm{d}k}\,\left\{a^{(r)}(k)\mathbf{f}^{(r)}(\mathbf{k})\exp(-\mathrm{i}kx)-a^{(r)\,\dagger}(k)\mathbf{f}^{(r)}(\mathbf{k})\exp(\mathrm{i}kx)\right\}\,,
\end{align}
\begin{equation}
\mathbf{B}=\boldsymbol{\nabla}\times \mathbf{A}=\sum_{r=1,2}\int \widetilde{\mathrm{d}k}\,\left\{a^{(r)}(k)\mathbf{b}^{(r)}(\mathbf{k})\exp(-\mathrm{i}kx)-a^{(r)\,\dagger}(k)\mathbf{b}^{(r)}(\mathbf{k})\exp(\mathrm{i}kx)\right\}\,,
\end{equation}
\end{subequations}
with the vector coefficients
\begin{subequations}
\begin{align}
\mathbf{f}^{(r)}(\mathbf{k})&\equiv \mathbf{g}^{(r)}(\mathbf{k})-\mathbf{h}^{(r)}(\mathbf{k})\,,\quad \mathbf{g}^{(r)}(\mathbf{k})\equiv \mathrm{i}\omega(\mathbf{k})\boldsymbol{\varepsilon}^{(r)}(\mathbf{k})\,,\quad \mathbf{h}^{(r)}(\mathbf{k})\equiv \mathrm{i}\mathbf{k}\varepsilon^{(r)\,0}(\mathbf{k})\,, \\
\mathbf{b}^{(r)}(\mathbf{k})&\equiv \mathrm{i}\mathbf{k}\times \boldsymbol{\varepsilon}^{(r)}(\mathbf{k})\,.
\end{align}
\end{subequations}
With these ingredients the expectation value of a bilinear combination of the electric field strength components, which are integrated over
configuration space, is computed. To simplify the notation, the spatial indices are set as lower ones:
\begin{align}
\label{eq:electric-field-bilinear}
E_{ij}&\equiv \langle \mathbf{k},\sigma|\int \mathrm{d}^3x\,:E_i(x)E_j(x):|\mathbf{k},\sigma\rangle \notag \displaybreak[0]\\
&=\sum_{r,s=1,2} \langle \mathbf{k},\sigma|\int \mathrm{d}^3x\,\widetilde{\mathrm{d}k}'\,\widetilde{\mathrm{d}k}''\,:\left[a^{(r)}(k')f_i^{(r)}(\mathbf{k}')\exp(-\mathrm{i}k'x)-a^{(r)\,\dagger}(k')f_i^{(r)}(\mathbf{k}')\exp(\mathrm{i}k'x)\right] \notag \\
&\phantom{{}={}\sum_{r,s=1,2} \langle \mathbf{k},\sigma|}\times \left[a^{(s)}(k'')f_j^{(s)}(\mathbf{k}'')\exp(-\mathrm{i}k''x)-a^{(s)\,\dagger}(k'')f_j^{(s)}(\mathbf{k}'')\exp(\mathrm{i}k''x)\right]:|\mathbf{k},\sigma\rangle \notag \displaybreak[0]\\
&=-\sum_{r,s=1,2} \langle \mathbf{k},\sigma|\int \widetilde{\mathrm{d}k}'\,:\frac{1}{2\omega(\mathbf{k}')}\left[a^{(r)}(k')a^{(s)\,\dagger}(k')f_i^{(r)}(\mathbf{k}')f_j^{(s)}(\mathbf{k}')\right. \notag \\
&\phantom{{}={}-\sum_{r,s=1,2} \langle \mathbf{k},\sigma|\int \widetilde{\mathrm{d}k}'\,:\frac{1}{2\omega(\mathbf{k}')}\Big[}\left.+\,a^{(r)\,\dagger}(k')a^{(s)}(k')f_i^{(r)}(\mathbf{k}')f_j^{(s)}(\mathbf{k}')\right]:|\mathbf{k},\sigma\rangle
\end{align}
Normal ordering moves all creation operators to the left, hence $:a^{(r)}(k')a^{(s)\,\dagger}(k'):=a^{(s)\,\dagger}(k')a^{(r)}(k')$.
Using $\langle \mathbf{k},\sigma|a^{(s)\,\dagger}(k')a^{(r)}(k')|\mathbf{k},\sigma\rangle=(2\pi)^32\omega(\mathbf{k})\delta(\mathbf{k}-\mathbf{k}')\delta_{r\sigma}\delta_{s\sigma}$
leads to:
\begin{equation}
E_{ij}=-\frac{1}{\omega(\mathbf{k})}f_i^{(\sigma)}(\mathbf{k})f_j^{(\sigma)}(\mathbf{k})\,.
\end{equation}
An analogous calculation for the $\mathbf{B}$-field with $f_i$ replaced by $b_i$ results in:
\begin{equation}
\label{eq:magnetic-field-bilinear}
B_{ij}\equiv \langle\mathbf{k},\sigma|\int \mathrm{d}^3x\,:B_i(x)B_j(x):|\mathbf{k},\sigma\rangle=-\frac{1}{\omega(\mathbf{k})}b_i^{(\sigma)}(\mathbf{k})b_j^{(\sigma)}(\mathbf{k})\,.
\end{equation}
Now Eqs.~(\ref{eq:electric-field-bilinear}), (\ref{eq:magnetic-field-bilinear}) can be used to obtain the normalization of the polarization
vectors by inserting these expressions into Eqs.~(\ref{eq:normalization-condition}), (\ref{eq:energy-momentum-tensor}).

\section{Computation of the gauge field commutator}
\label{sec:gauge-field-commutator-calculation}

The commutator $K^{\mu\nu}(x)$ of two vector potentials (one evaluated at a generic spacetime point $x_1=x$ and the other at $x_2=0$) can be
calculated with \eqref{eq:vector-potential}. It is written as an integral over a commutator $\widehat{K}^{\mu\nu}$ in momentum space:
\begin{subequations}
\begin{equation}
K^{\mu\nu}(x)\equiv [A^{\mu}(x),A^{\nu}(0)]=\sum_{r,s=1,2} \int \widetilde{\mathrm{d}k}\,\widetilde{\mathrm{d}k}'\,\widehat{K}^{\mu\nu}(k,k')\,,
\end{equation}
with
\begin{align}
\widehat{K}^{\mu\nu}(k,k')&=[K^{\mu}(k),K^{\nu}(k')]\,, \\[2ex]
K^{\mu}(k)&=a^{(r)}(k)\varepsilon^{(r)\,\mu}(\mathbf{k})\exp(-\mathrm{i}kx)+a^{(r)\,\dagger}(k)\overline{\varepsilon}^{(r)\,\mu}(\mathbf{k})\exp(\mathrm{i}kx)\,, \\[2ex]
K^{\nu}(k')&=a^{(s)}(k')\varepsilon^{(s)\,\nu}(\mathbf{k}')+a^{(s)\,\dagger}(k')\overline{\varepsilon}^{(s)\,\nu}(\mathbf{k}')\,.
\end{align}
\end{subequations}
Now the evaluation of the aforementioned commutator in momentum space yields:
\begin{align}
\widehat{K}^{\mu\nu}(k,k')&=[a^{(r)}(k),a^{(s)\,\dagger}(k')]\varepsilon^{(r)\,\mu}(\mathbf{k})\overline{\varepsilon}^{(s)\,\nu}(\mathbf{k}')\exp(-\mathrm{i}kx) \notag \\
&\phantom{{}={}}+[a^{(r)\,\dagger}(k),a^{(s)}(k')]\overline{\varepsilon}^{(r)\,\mu}(\mathbf{k})\varepsilon^{(s)\,\nu}(\mathbf{k}')\exp(\mathrm{i}kx) \notag \\
&=(2\pi)^32\omega(\mathbf{k})\delta_{rs}\delta^{(3)}(\mathbf{k}-\mathbf{k}')\big[\varepsilon^{(r)\,\mu}(\mathbf{k})\overline{\varepsilon}^{(s)\,\nu}(\mathbf{k}')\exp(-\mathrm{i}kx) \notag \\
&\phantom{{}={}(2\pi)^32\omega(\mathbf{k})\delta_{rs}\delta^{(3)}(\mathbf{k}-\mathbf{k}')\big(}-\overline{\varepsilon}^{(r)\,\mu}(\mathbf{k})\varepsilon^{(s)\,\nu}(\mathbf{k}')\exp(\mathrm{i}kx)\big] \notag \\
&=(2\pi)^32\omega(\mathbf{k})\delta_{rs}\delta^{(3)}(\mathbf{k}-\mathbf{k}')\,\varepsilon^{(r)\,\mu}(\mathbf{k})\varepsilon^{(s)\,\nu}(\mathbf{k})\left[\exp(-\mathrm{i}kx)-\exp(\mathrm{i}kx)\right]\,.
\end{align}
In the last step it was used that the corresponding polarization vectors were chosen as real in \secref{eq:polarization-vectors}.
Using the definition of the polarization sum $\Pi^{\mu\nu}$ of \eqref{eq:polarization-sum} the commutator in configuration space can be
obtained as
\begin{subequations}
\begin{align}
K^{\mu\nu}(x)&=\int \frac{\mathrm{d}^3k}{(2\pi)^32\omega(\mathbf{k})}\,\Pi^{\mu\nu}(k)\left[\exp(-\mathrm{i}kx)-\exp(\mathrm{i}kx)\right] \notag \displaybreak[0]\\
&=\int \frac{\mathrm{d}^3k}{(2\pi)^3}\,\Pi^{\mu\nu}(k)\left[\left.\frac{\exp(-\mathrm{i}kx)}{2\omega(\mathbf{k})}\right|_{k^0=\omega(\mathbf{k})}+\left.\frac{\exp(-\mathrm{i}kx)}{-2\omega(\mathbf{k})}\right|_{k^0=-\omega(\mathbf{k})}\right] \notag \displaybreak[0]\\
&=-\mathrm{i}\oint_C \frac{\mathrm{d}k^0}{2\pi} \int \frac{\mathrm{d}^3k}{(2\pi)^3}\,\Pi^{\mu\nu}(k)B(k)\exp(-\mathrm{i}kx) \notag \displaybreak[0]\\
&=-\mathrm{i}\oint_C \frac{\mathrm{d}^4k}{(2\pi)^4}\,\Pi^{\mu\nu}(k)B(k)\exp(-\mathrm{i}kx)\,. \displaybreak[0]\\[2ex]
B(k)&=\frac{1}{[k^0-\omega(\mathbf{k})][k^0+\omega(\mathbf{k})]}\,.
\end{align}
\end{subequations}
Here the contour $C$ is chosen such that it encircles all poles of $B(k)$ in counterclockwise direction. Finally the normalization of the polarization
vectors is put into $B(k)$ to get the scalar propagator part $\widehat{K}$. The result then reads
\begin{subequations}
\label{eq:commutator-vector-potentials}
\begin{align}
K^{\mu\nu}(x)&=\mathrm{i}\theta^{\mu\nu}(\mathrm{i}\partial^{\mu},\mathrm{i}\partial^{\nu})\int_C \frac{\mathrm{d}^4k}{(2\pi)^4}\,\widehat{K}\exp(-\mathrm{i}kx)\,, \\[2ex]
\theta^{\mu\nu}(k^{\mu},k^{\nu})&=\eta^{\mu\nu}+\frac{1}{\mathbf{k}^2}\,k^{\mu}k^{\nu}-\frac{k^0}{\mathbf{k}^2}\,(k^{\mu}\xi^{\nu}+\xi^{\mu}k^{\nu})-\left(1-\frac{(k^0)^2}{\mathbf{k}^2}\right)\,\xi^{\mu}\xi^{\nu}\,.
\end{align}
\end{subequations}
Here the second-rank tensor $\theta^{\mu\nu}$ is expressed via derivatives to reproduce the structure of the polarization tensor $\Pi^{\mu\nu}$
when acting on the complex exponential function. This shows that the scalar properties of the commutator are encoded in the four-dimensional
contour integral over the scalar propagator part $\widehat{K}$.

Note that this derivation is only valid for the spatial sector. That is why in \secref{sec:gauge-field-commutator} a special case of the
spatial sector is considered. The computation of the scalar commutator function has to be modified for the temporal and the mixed sectors.
This is not within the scope of the current article, though, but it is an interesting open problem for future studies.

\end{appendix}

\newpage%%tmp
%---------------------------------------------------------------------------------------------------

%--------------------------------------------------------------------------------------------------

\end{document}